\documentstyle[aaspp4]{article}
\def \etal {{\it et al. }}

\slugcomment{To appear in Ap.J.}

\lefthead{}
\righthead{}
\def\deg{\mbox{$^\circ$}}

\def\littleprime{\ifmmode{\scriptscriptstyle \prime } 
\else{\hbox{$\scriptscriptstyle \prime$ }}\fi}

\def\arcsec{\raise .9ex \hbox{\littleprime\hskip-3pt\littleprime}}
\def\arcmin{\raise .9ex \hbox{\littleprime}}
\def\arcsecpoint{\hbox to 1pt{}\rlap{\arcsec}.\hbox to 2pt{}}
\def\arcminpoint{\hbox to 1pt{}\rlap{\arcmin}.\hbox to 2pt{}}

\begin{document}

\title{Quantitative Morphology of Galaxies in the Hubble Deep Field}

\author{Francine R. Marleau}
\affil{Department of Astronomy, University of California, Berkeley, \\
Campbell Hall, Berkeley, CA 94720, USA}
\author{and} 
\author{Luc Simard}
\affil{UCO/Lick Observatory, University of California, Santa Cruz, CA 
95064, USA}

\begin{abstract}
We measure quantitative structural parameters of galaxies in the 
Hubble Deep Field (HDF) on the drizzled F814W images.  Our structural 
parameters are based on a two-component surface brightness made up of 
a S\'ersic profile and an exponential profile.  We compare our results 
to the visual classification of van den Bergh \etal (1996) and the 
$C-A$ classification of Abraham \etal (1996a).  Our morphological 
analysis of the galaxies in the HDF indicates that the spheroidal 
galaxies, defined here as galaxies with a dominant bulge profile, make 
up for only a small fraction, namely 8\% of the galaxy population down 
to m$_{F814W}(AB)$ = 26.0.  We show that the larger fraction of 
early-type systems in the van den Bergh sample is primarily due to the 
difference in classification of 40\% of small round galaxies with 
half-light radii $<$ 0\arcsecpoint 31.  Although these objects are
visually classified as elliptical galaxies, we find that 
they are disk-dominated with bulge fractions $<$ 0.5.
Given the existing large dataset of HDF galaxies with measured 
spectroscopic redshifts, we are able to determine that the majority of 
distant galaxies ($z>2$) from this sample are disk-dominated.
Our analysis reveals a subset of HDF 
galaxies which have profiles flatter than a pure exponential profile.
\end{abstract}


\newpage

\section{Introduction} \label{intro}
Starting from our knowledge of gravity and the effects of local 
gravitational instabilities on the evolution of the Universe, we can 
infer scenarios for galaxy formation that predict the current state of 
the Universe.  The evolutionary path followed by these perturbations 
as they develop to become present-day galaxies is poorly constrained.  
It depends on the star formation history of galaxies and the initial 
conditions of the gravitational collapse (Steinmetz \& M\"uller 1995), 
and both are not directly observable.  However, the structure and 
kinematics of galaxies, which are tied to their formation process 
(Okamura \etal 1988), are quantities that can be measured from 
observations.  The structural parameters (shape, size, axial ratios, 
etc.)  of large samples of distant galaxies play therefore an 
important if not unique role in the multi-parameter study that is 
inevitably required to understand and explain the origin of galaxies.

The properties of the galaxy population over the redshift range 
$0<z<1$ have been studied extensively from faint redshift surveys 
(Lilly \etal 1996; Cowie \etal 1996; Ellis \etal 1996).  These surveys 
have shown that the galaxy population is evolving in number density 
and/or luminosity and that this evolution depends strongly on color.  
However, galaxy number counts and redshifts are insufficient to 
determine which part of the galaxy population is evolving and how.  In 
order to understand the role of each galaxy type in the evolution seen 
at intermediate redshift, morphological information must be extracted 
for field galaxies using an objective classification method.  For the 
nearby sample, the existing morphological classification is based 
solely on visual inspection of galaxy $B$-band images, a method 
introduced with the work of Hubble (1926, 1930) and developed by de 
Vaucouleurs (1959) and Sandage (1961).  The surface brightness profile 
of galaxies can in fact be quantified using empirical luminosity laws 
such as the $r^{1/4}$ law associated with the spheroidal populations 
(de Vaucouleurs 1948, 1953) and the exponential profile describing the 
disks of galaxies (Patterson 1940; de Vaucouleurs 1956).  These 
profiles exist for a restricted number of nearby field and cluster 
galaxies and have not been used until recently, with the development 
of photometric decomposition methods (Schade \etal 1996; Abraham \etal 
1996b), to classify distant galaxies in a consistent way.

Visual classification is increasingly difficult for faint or high 
redshift galaxies, or both, and it is therefore necessary to use a quantitative 
profile decomposition method to retrieve the physical properties of 
the observed two-dimensional (2D) images of galaxies.  When comparing 
nearby galaxy samples with high redshift galaxies, it is also 
essential to establish a correspondence between visual and 
quantitative classifications.  Quantitative classification has two 
major advantages over visual classification: (1) it is reproducible, 
and (2) biases can be understood and carefully characterized through 
simulations which are treated as real data.  The detailed images of 
the Hubble Deep Field (HDF) provide an ideal dataset of nearby and 
distant field galaxies for the study of morphological properties of 
galaxies in the context of evolution.  The HDF is a Hubble Space 
Telescope (HST) program that has imaged a field in the northern 
continuous viewing zone for ten consecutive days, or approximately 150 
orbits, in four passbands (Williams \etal 1996).  The morphologies of 
HDF galaxies in the range $21< I <25$ have been visually classified 
(van den Bergh \etal 1996, hereafter VDB96) and measured using a 
quantitative classification system based on the study of the central 
concentration and asymmetry of the galaxian light (Abraham \etal 
1996a, hereafter ABR96).  These classification techniques find that the fraction 
of elliptical galaxies in the HDF is as large as 30\% (although the 
classification by ABR96 gives a value closer to 20\%), with the
remainder being divided into 31\% spirals and 39\% unclassified. 

In this paper, we examine the structural properties of the galaxies in 
the HDF using a new 2D photometric decomposition fitting algorithm.  
We present the structural parameter distributions of HDF galaxies and 
compare our results to previous classification schemes.  The outline 
of this paper is as follows: Section~\ref{2dmodel} describes the 
photometric decomposition method that we use to determine 
morphological properties of the HDF galaxies.  In this section, we 
also present the results of simulations that test the ability of our 
method to measure reliably galaxian structural parameters.  In 
Section~\ref{results} we present the results of our photometric 
decomposition technique applied to the HDF. Section~\ref{comparison} 
considers previous classification schemes of HDF galaxies and 
underlines the differences between classification methods.  
Section~\ref{profano} deals with the limitations of the standard 
bulge/disk decomposition model and the observed deviations from that 
model for a number of galaxies in the HDF. We summarize our 
conclusions and discuss their implications for galaxy evolution in 
Section~\ref{discuss}.

\section{Two-Dimensional Modeling of Galaxy Images} \label{2dmodel}

\subsection{Photometry} \label{photom}
We used the SExtractor galaxy photometry package version 1.0a (Bertin 
\& Arnouts 1996) on the publicly released version 2 drizzled $F814W$ 
images of the HDF with a detection threshold of 1.5$\sigma$ and a 
minimum object area of 30 pixels.  The total sky area analyzed was 
16085 arcsec$^{2}$.  The resulting SExtractor catalogs contained 566 
objects from WFPC2-chip 2, 498 objects from chip 3 and 575 objects 
from chip 4 for a total of 1639 objects.  As SExtractor performs 
galaxy photometry, it constructs a ``segmentation'' image.  Pixels 
belonging to the same object all have the same values in this 
segmentation image.  It can therefore be used as a pixel mask for the 
surface brightness profile fits.  SExtractor deblends objects using 
multiple flux thresholding.  At each flux threshold, it computes the 
number of possible independent ``branches'', and the fraction of the 
total flux contained in each one.  The SExtractor deblending parameter 
{\sl DEBLEND$_-$MINCONT} sets the minimum fraction of the total flux a 
branch must contain to be considered as a separate object.  We used a 
{\sl DEBLEND$_-$MINCONT} value of 0.001 as our definition of a 
distinct object.  As shown in Section~\ref{pardist}, our galaxy number 
counts are identical to those of previous investigators (e.g.  Bouwens 
\etal 1997).

For each object, we used the following SExtractor parameters: 
centroid $X$ and $Y$, local sky background level and variance and the 
1.5$\sigma$ isophotal area.  We extracted an area around each object 
20 times larger than its isophotal area from both the science image 
and the segmentation image to ensure that our fitting routine would 
successfully discriminate between galaxy and sky fluxes.

\subsection{Surface Brightness Profile Model} \label{surfbmodel}
The structure of each detected galaxy in the HDF was examined carefully
using GIM2D (Galaxy IMage 2D), a 2D decomposition fitting program
(Simard 1998).  GIM2D is an IRAF/SPP package written to perform detailed
surface brightness profile decompositions of low signal-to-noise (S/N)
images of distant galaxies in a fully automated way.  GIM2D takes an
input image, does a 2D profile fit on the image pixels belonging to the
same pixel value segmentation image, and produces a galaxy-subtracted
image as well as a catalog of structural parameters.  The 2D galaxy
model used by GIM2D has a maximum of twelve parameters: the total flux
$F$ in data units (DU), the bulge fraction $B/T$ (0=pure disk system),
the bulge effective radius $r_e$, the bulge ellipticity $e$ ($e \equiv
1-b/a$, $b \equiv$ semi-minor axis, $a \equiv$ semi-major axis), the
bulge position angle of the major axis $\phi_{b}$ (clockwise, 0=y-axis),
the exponential disk scale length $r_d$, the disk inclination $i$
(0=face-on), the disk position angle $\phi_d$, the subpixel $dx$ and
$dy$ offsets of the galaxy center, the background level $b$, and the
S\'ersic index $n$.  One or more parameters can be frozen to some
initial values if necessary.  We did not constrain $\phi_b$ and $\phi_d$
to be equal for two reasons: (1) a large difference between these
position angles is a signature of barred spirals, and (2) we have
observed galaxies with {\it bona fide} bulges which were not quite
aligned with the disk position angle.  All total fluxes $F$ were
converted in this paper to $F814W$ magnitudes on the AB system using the
equation:
\begin{equation}
	m_{F814W}(AB) = -2.5 log_{10}(F/t) + C, 
	\label{magdef}
\end{equation}
\noindent where $C$ = 22.09 for WF2, 22.09 for WF3, and 22.07 for WF4.  
The total exposure time $t$ was 123600 seconds in the $F814W$ filter 
(Williams \etal 1996).

The first component (``bulge'') of the 2D surface brightness used by 
GIM2D to model galaxy images is a S\'ersic (1968) profile of the form:
\begin{equation}
	\Sigma(r) = \Sigma_{e} exp \{-b[(r/r_{e})^{1/n} - 1]\}, 
 	\label{sersic}
\end{equation} 
\noindent where $\Sigma(r)$ is the surface brightness at radius $r$.  The 
parameter $b$ is set equal to 1.9992$n-$0.3271 so that $r_{e}$ 
remains the projected radius enclosing half of the light in this 
component (Capaccioli 1989).  The classical de Vaucouleurs profile 
therefore has the special value $n$ = 4.  The second component 
(``disk'') is a simple exponential profile of the form:
\begin{equation}
	\Sigma(r) = \Sigma_{0} exp (-r/r_d). 
	\label{disk}
\end{equation}
\noindent $\Sigma_{0}$ is the central surface brightness.  
We adopted here the conventional ``bulge/disk'' nomenclature and assumed 
this distinction in the galaxies light profile classification throughout
this paper.  Nevertheless, it should be kept in mind that this 
nomenclature does not say anything 
about the internal kinematics of the components.  The presence of a 
``disk'' component does not imply the presence of an actual disk since 
many virially-supported systems also have simple exponential profiles.

The WFPC2 detector undersampling was taken into account by generating 
the surface brightness model on an oversampled grid, convolving it 
with the appropriate point spread function (PSF), shifting its center 
according to $dx$ and $dy$ and rebinning the result to the detector 
resolution for direct comparison with the observed galaxy image.  The 
PSF was generated by the Space Telescope package {\sl TINY TIM} (Krist 
1993) and subsampled to reproduce the pixel resolution of the HDF. As 
a first pass for our morphological analysis, we fitted all the objects 
in our HDF catalog as the sum of a de Vaucouleurs profile and a simple 
exponential.  However, as discussed in Section~\ref{profano}, we 
discovered that this model failed for a number of galaxies which had 
surface brightness profiles {\it flatter} than a {\it pure} ($B/T=0$) 
exponential profile.  The second pass in our morphological analysis 
therefore consisted of fitting {\it pure} S\'ersic profiles to these 
galaxies with a S\'ersic index $n$ allowed to vary between 0.2 and 
4.0.

\subsection{Fitting Algorithm} \label{fitalgo}
The 12-dimensional parameter space can have a very complicated 
topology with local minima at low S/N ratios.  It was therefore 
important to choose an algorithm which did not easily get fooled by 
those local minima.  The Metropolis algorithm (Metropolis \etal 1953, 
Saha \& Williams 1994) was designed to search for optimal parameter 
values in a complicated topology.  Compared to gradient search 
methods, the Metropolis is not efficient i.e.  it is CPU intensive.  
On the other hand, gradient searches are lazy.  They will start from 
initial parameter values, dive in the first minimum they encounter and 
claim it is the global one.

The Metropolis algorithm in GIM2D starts from an initial set of 
parameters given by the image moments of the object and computes the 
likelihood $P(w|D,M)$ that the parameter set $w$ is the true one given 
the data $D$ and the model $M$.  It then generates random 
perturbations $\Delta{\bf x}$ about that initial location with a given 
``temperature''.  When the search is ``hot'', large perturbations are 
tried.  After each trial perturbation, the Metropolis algorithm 
computes the likelihood value $P_{1}$ at the new location, and 
immediately accepts the trial perturbation if $P_{1}$ is greater than 
the old value $P_{0}$.  However, if $P_{1}$ $<$ $P_{0}$, then the 
Metropolis algorithm will accept the trial perturbation only 
$P_{1}$/$P_{0}$ of the time.  Therefore, the Metropolis algorithm will 
sometime accept trial perturbations which take it to regions of lower 
likelihood.  This apparently strange behavior is very valuable: if the 
Metropolis algorithm finds a minimum, it will try to get out of it, 
but it will only have a finite probability (related to the depth of 
the minimum) of succeeding.  The ``temperature'' is regulated 
according to the number of accepted iterations.  If the Metropolis 
accepts too many trial perturbations, then the search is too ``cold'', 
and the temperature must be increased.  Conversely, if the Metropolis 
rejects too many trial perturbations, then the search is too ``hot'', 
and the temperature must be decreased.  The Metropolis temperature is 
regulated such that half of the trial perturbations are accepted.  The 
more commonly known simulated annealing technique is a variant and 
a special case of the Metropolis algorithm in which the temperature is 
only allowed to decrease until the ``ground-state'' is reached.

The step matrix for the trial perturbations $\Delta{\bf x}$ is given 
by the simple equation $\Delta{\bf x} = Q \cdot \vec{u}$ where the vector 
{\bf$\vec{u}$} consists of randomly generated numbers between 0 and 1, and 
the matrix Q is obtained through Choleski inversion of the local 
covariance matrix of accepted iterations (Vanderbilt \& Louie 1984).  
In short, the sampling of parameter space shapes itself to the local 
topology.

Convergence is achieved when the difference between two likelihood 
values separated by 100 iterations is less than 3$\sigma$ of the 
likelihood fluctuations.  After convergence, the Metropolis algorithm 
Monte-Carlo samples the region where the likelihood is thus maximized 
and stores the accepted parameter sets as it goes along to build the 
distribution $P(w|D,M)$.  Once the region has been sufficiently 
sampled, the Metropolis algorithm computes the median of $P(w|D,M)$ 
for each model parameter as well as the 99\% confidence limits.  The
output of the fitting process consists of a PSF-convolved model image 
$O$, a residual image $R$, and a log file containing all Metropolis 
algorithm iterations, the final parameter values and their confidence 
intervals.  

\subsection{Asymmetry Index} \label{asymind}
The residuals from the smooth model fits were analyzed to give 
further information on the morphology of the galaxies.  The nature of 
asymmetric residuals is of particular interest to galaxy evolution.  
For example, the presence of star-forming regions or a recent merger 
event generally give rise to asymmetric features.  An asymmetry 
index was extracted from the residual image to assess the nature of 
the non-smooth 2D profiles.  For each object analyzed by GIM2D, a 
reduced chi-square $\chi^2_R$, a measure of the residual flux,
and a seeing-deconvolved half-light radius $r_{hl}$ 
were obtained.  The half-light radius was computed by integrating 
equations~\ref{sersic} and~\ref{disk} with the measured structural 
parameters.  The parameter $R_A$ was calculated using the 
following expression:
\begin{equation} 
	R_A = \frac{\frac{1}{2} \Sigma |R_{ij} - R_{ij}^{180}|}{\Sigma I_{ij}}, 
\label{rart}
\end{equation}
where $I_{ij}$, $R_{ij}$ and $R_{ij}^{180}$ are the flux at ($i$,$j$) 
in the input original image, the residual image, and the residual 
image rotated by 180 degrees, respectively.  Aperture sizes ranging 
from 1$-$10 $r_{hl}$ were used.  The index $R_A$ is calculated over 
the same pixels used in the bulge/disk fits.  This parameter is 
equivalent to the asymmetric residual flux index defined in 
Schade \etal (1995) when computed within a radius of 5 $h_{50}^{-1}$ 
kpc.  Background noise absolute pixel values in equation~\ref{rart} 
can make significant contributions to $R_A$ even in the 
absence of any residual.  This background contribution was removed by 
computing a correction for $R_A$ over a sky background area equal 
to the object area and subtracting the result from the ``raw''  
$R_A$ values.  The background corrected $R_A$ converged to 
constant values at large radii.

\section{Results of the Photometric Decompositions} \label{results}

\subsection{The HDF Catalog} \label{hdfcat}
Profile fitting was done for a total of 1639 objects in the HDF. For 
each galaxy, we obtained a PSF-convolved model image, a residual 
image, the best parameter values with their confidence limits, and the 
reduced chi-square $\chi^2_R$.  Figure~\ref{hd4image} shows a section of 
the WFPC2-chip 4 HDF image and the residual image created by GIM2D 
after detailed photometric decompositions have been performed on all 
galaxies in the field.

In Table~1 we present the best structural parameter values with their 
confidence limits and the goodness of the fit for the 522 galaxies 
with $m_{F814W} \leq 26.0$ in the HDF analyzed with GIM2D. The 
S\'ersic parameters for the subsample of HDF galaxies discussed in 
Section~\ref{profano} are given in Table~2.  Only the WFPC2-chip 4 
sample of these extensive tabulations of data are printed here and 
the complete tables are stored in the electronic archive of 
The Astrophysical Journal, from which they can be retrieved with 
the standard procedure (also available at 
http://astro.berkeley.edu/$\sim$marleau/).

\subsection{Bulge and Disk Structural Parameter Distributions} 
\label{pardist}
The HDF is the deepest field ever imaged with the superior angular 
resolution of HST. Consequently, the structural parameter 
distributions of HDF galaxies provide unique tests of galaxy evolution 
models (Bouwens {\it et al.} 1997). This section focuses on the {\it 
observed} and {\it intrinsic} structural parameter distributions of HDF 
galaxies. Intrinsic distributions fully take selection effects into 
account and should be compared with theoretical predictions.

The {\it observed} distributions of structural parameters for our 
entire sample of 1639 galaxies are shown in 
Figure~\ref{observed-dist-all}.  Bulge parameter distributions 
($r_{e}$, $e$, $\phi_b$) only include objects with $B/T>0.5$ whereas 
disk parameter distributions ($r_d$, $i$, $\phi_d$) only include 
objects with $B/T \leq 0.5$.  Such a separation is needed to minimize 
the scatter in those distributions caused by poorly constrained bulge 
parameters in disk-dominated galaxies and vice versa.  
Figure~\ref{observed-dist-tmlt26} shows the same distributions for 
galaxies brighter than $m_{F814W}(AB) = 26.0$.  The distributions of 
half-light radii for different $m_{F814W}(AB)$ cuts are shown in 
Figure~\ref{observed-dist-hlr}.  The half-light radius versus $B/T$ 
distribution in Figure~\ref{observed-dist-hlr} suggests that there are 
no large ellipticals in our HDF sample.

The observed parameter distributions of galaxies must be corrected for 
the galaxy selection function of the SExtractor detection algorithm to 
produce the final intrinsic parameter distributions.  The detection 
thresholding method used by SExtractor depends critically on galaxy 
apparent surface brightness.  The probability that a given object will 
be detected depends on total flux $F$, bulge fraction $B/T$, bulge 
effective radius $r_e$, bulge ellipticity $e$, disk scale length 
$r_d$, and disk inclination $i$.  For example, objects with larger 
$B/T$ will be easier to detect because they are more concentrated, and 
large objects will be harder to detect than smaller ones at a fixed 
total flux.  In order to derive true intrinsic structural parameter 
distributions, we must therefore construct a six-dimensional galaxy 
selection function $S(\omega)$ over all parameter space locations 
$\omega \equiv (F,B/T,r_e,e,r_d,i)$ and use it to ``flat-field'' the 
observed structural parameter distributions.  The selection function 
does not depend on the bulge and disk position angles.  

We built $S(\omega)$ by generating 66000 galaxies that were modeled 
with structural parameter values uniformly covering the structural 
parameter ranges: $23.2 \leq m_{F814W}(AB) \leq 29.0$, $0 \leq B/T 
\leq 1$, $0 \leq r_{e} \leq 0\arcsecpoint 8$, $0 \leq e \leq 0.7$, $0 
\leq r_{d} \leq 0\arcsecpoint 8$, $0 \leq i \leq 85\deg$.  Each model 
galaxy was added one at a time to an empty section of the HDF covering 
6\arcsecpoint0$\times$5\arcsecpoint 2.  ``Empty'' here means that no 
objects were detected by SExtractor in that sky section with the same 
detection parameters used to construct our object catalog.  Using an 
empty section of the HDF ensured that we were building $S(\omega)$ 
with the real background noise that was ``seen'' by the detection 
algorithm.  The background ``noise'' included read-out, sky and the 
brightness fluctuations of all the very faint galaxies that were 
beyond our detection threshold.  This last contribution to the 
background noise is particularly hard to model theoretically, and our 
approach bypassed this problem.  We divided parameter space into 320 
``cells'' with the number of cells along each dimensions being 
($N_{m},N_{B/T},N_{r_{e}},N_{e},N_{r_{d}},N_{i}, N_{r_{hl}}$) = 
(8,5,1,1,1,1,8).  SExtractor was run on all 66000 images to determine 
the number of detected models $D(\omega)$ in each cell.  The selection 
function $S(\omega)$ (shown in Figure~\ref{selfplot}) was simply $D(\omega)$ 
divided by the number of models $G(\omega)$ generated in that cell in 
parameter space.  $S(\omega)$ was set automatically to one for cells 
with $m_{F814W}(AB) \leq 23.2$.  One-dimensional distributions for 
$m_{F814W}(AB)$, $B/T$, and $r_{hl}$ were created by integrating over 
all cells with $S(\omega)$ greater than or equal to a fixed limit 
$S_{lim}$.

Even though measured structural parameters were not reliable fainter 
than $m_{F814W}(AB) = 26.0$, the observed and intrinsic distributions 
of $m_{F814W}(AB)$, $B/T$, and $r_{hl}$ over the magnitude range $21.0 
\leq m_{F814W}(AB) \leq 29.0$ (Figure~\ref{intrinsic-dist-2129}) were 
first calculated to provide an ``upper limit'' on the intrinsic parameter 
distributions.  The top and bottom distributions are for $S_{lim} \geq 
0.1$ and $S_{lim} \geq 0.5$.  Figure~\ref{intrinsic-dist-2126} 
displays the same distributions as Figure~\ref{intrinsic-dist-2129} 
for $21.0 \leq m_{F814W}(AB) \leq 26.0$.  Except for the case where 
$21.0 \leq m_{F814W}(AB) \leq 29.0$ and $S_{lim}=0.1$, the selection 
function corrections are relatively small.  It is worth noting that 
all $B/T$ intrinsic distributions show that the detection algorithm is 
biased against pure disk systems (see the $21.0 \leq m_{F814W}(AB) 
\leq 26.0$ distributions for example) as expected since pure disk 
systems are less centrally concentrated than $r^{1/4}$ profiles and 
thus harder to detect.  So, the intrinsic ratio of the number of 
disk-dominated systems to the number of bulge-dominated is likely to 
be higher than the observed one.  

\subsection{Asymmetry Parameter Distribution} \label{asympardist}
The deviation from smoothness and symmetry are striking 
features of the morphologies of HDF galaxy images.  Some fraction 
of the galaxies in the HDF contain profile irregularities which may 
significantly perturb fits of elliptically symmetric two-dimensional 
models.  These irregularities have two causes: (1) the morphologies of 
high redshift spiral and starburst galaxies change radically as the 
observed bandpass shifts to the rest-frame UV where HII regions dominate 
the galaxy light distribution (Giavalisco \etal 1996a), and (2) some 
galaxies are just intrinsically more disturbed.  We analyzed the residuals 
of our galaxy sample with $m_{F814W}(AB) \leq 26.0$ by evaluating the 
asymmetry index given in Section~\ref{asymind}.  The results are 
presented in Figure~\ref{ra12} for $R_A$ computed within a 2$r_{hl}$ 
aperture.  Also shown in Figure~\ref{ra12} is a measure of the residual flux 
$\chi^2_R$ as a function of the degree of asymmetry $R_A$ for our sample 
of galaxies.  The index $R_A$ remains smaller
than $\sim$20\% for the majority of galaxies 
(those objects with $R_A>0.3$ are positioned near the edge of the chip 
where it is most difficult to evaluate the sky background correction  
accurately).  

Non-smooth local features in a galaxy 2D light profile can alter the 
best parameters derived with GIM2D depending on their brightnesses and 
positions in the galaxy.  For example, a very bright feature at the 
center of the galaxy will cause the bulge component to be 
overestimated.  We measured the effects of clumps or asymmetric 
features on the extracted smooth 2D profile parameters through 
simulations.  An asymmetric light component, in the form of one or 
multiple ``blobs'', i.e.  unresolved sources convolved with the PSF, 
was added to the simulated smooth 2D profile image.  The input parameters 
for generating the asymmetric features are the number of blobs 
$n_b$, the fraction of total galaxy flux in the blobs $F_b$, and the 
blobs' maximum galactocentric distance $r_b$.  The positions and 
fluxes of the blobs were randomly generated within the limits imposed 
by $F_b$ and $r_b$.

Asymmetric features superposed on the smooth profile were generated 
randomly for $n_b=5$ and $r_b=1.5r_{hl}$.  We tested five discrete 
flux levels ($F_b=$0.0, 0.05, 0.10, 0.15, 0.20) in order to sample the 
same range of residual fluxes as seen in the real data (see 
Figure~\ref{ra12}).  For each given $F_b$, we generated 10 galaxies 
with the following structural parameters: m$_{F814W}(AB)=24.0$, 
$B/T=0.3$, $r_{e}=0\arcsecpoint 12$, $e=0.2$, $r_{d}=0\arcsecpoint 
32$, $i=20\deg$, $\phi_b=\phi_d=60\deg$, and $n=4.0$.  The 50 
simulated asymmetric galaxy images were analyzed exactly the same way 
as the real data.  The detections were classified by SExtractor as 
``single'' or ``multiple''.  The average number of detections in a 
single simulated galaxy image increased with $F_b$ as expected.  As 
for the real data, profile fitting was done on the simulated image 
pixels belonging to the same pixel value segmentation image.  An 
example of a simulated image with a flux fraction $F_b=0.10$ is shown 
in Figure~\ref{simulb}.  For each $F_b$, we were able to examine the 
parameters recovered by GIM2D. The recovery success for the bulge 
fraction parameter are displayed in Figure~\ref{simulb}.  
The mean of the measured $B/T$ going from $F_b = 0.0$ to
$F_b = 0.20$ are 0.299 ($\sigma$=0.016), 0.270 (0.084), 0.253 (0.080),
0.282 (0.161), and 0.411 (0.218) with a total number of galaxies per bin of 
10.  The simulations showed that the perturbations to the profiles caused by 
the presence of asymmetric features did not systematically and 
significantly alter the measured bulge fraction over the 
range of galaxy asymmetric residual fluxes seen in the data.  The RMS 
(root-mean-square), which increased with $F_b$, ranged 
from 1-20\%, and the mean remained close to the input value of $B/T=0.3$.
Therefore, we do not expect the non-smooth galaxy features to change  
the total distribution of $B/T$ we derived from our galaxy sample in 
Section~\ref{pardist}.  

\subsection{Structural Parameter Recovery Simulations} 
\label{parrecov}
The results of this paper discussed below in Sections~\ref{comparison} 
and~\ref{profano} rely on our ability to measure (1) accurate 
structural bulge/disk parameters for small galaxies and (2) the index 
$n$ of pure S\'ersic profiles over the range 0.2 $\leq$ $n$ $\leq$ 4.  
We tested the reliability of our structural parameters through 
simulations.  We created 600 galaxies with structural parameters 
uniformly generated at random in the following ranges: m$_{F814W}(AB) 
\geq 24.4$, $0.0 \leq B/T \leq 1.0$, $0 \leq r_{e} \leq 0\arcsecpoint 
3$, $0 \leq e \leq 0.7$, $0 \leq r_{d} \leq 0\arcsecpoint 3$, $0 \leq 
i \leq 85\deg$, and $0.2 \leq n \leq 4.0$.  The $r_{e}$ and $r_d$ 
ranges cover the bulk of the real HDF galaxy size distributions (see 
Section~\ref{pardist}).  We did not impose any correlation between structural 
parameters.  Sky subtraction errors can lead to substantial errors on 
structural parameter estimates especially for steep profiles such as 
the $r^{1/4}$ law profile.  Indeed, the half-light radius of a 
bulge-dominated system can be underestimated if a significant fraction 
of the total flux is hidden in the outer wings of the profile where 
sky noise dominates.  To characterize the effects of {\it real} sky 
subtraction errors on structural parameter estimates, we used the 
empty section of the HDF described in Section~\ref{pardist}.  Our 600 simulated 
bulge/disk models were added to this background image one at a time 
and reduced as if they were real objects.

Our simulations showed that we systematically underestimated the total 
magnitudes of our objects by 0.1 mag over the magnitude range $24.4 
\leq m_{F814W}(AB) \leq 26.0$ with a dispersion of 0.05 mag.  The 
fractional total flux error quickly increased beyond 20$\%$ for 
objects fainter than $m_{F814W}(AB) = 26.0$.  Our structural 
parameters are therefore reliable for objects brighter than 
$m_{F814W}(AB) = 26.0$.  Out of the 1639 HDF objects originally 
analyzed with GIM2D, 522 satisfied this magnitude selection cut.

We classified HDF galaxies according to bulge fraction, and it was 
therefore very important to determine the reliability of our bulge 
fraction estimates down to our magnitude limit of $m_{F814W}(AB) = 
26.0$.  Figure~\ref{simul-results-hlr+bt} shows the difference between 
observed and input bulge fractions versus seeing-deconvolved 
half-light radius and $B/T$ for three different $B/T$ intervals ($0 
\leq B/T \leq 0.2$, $0.2 < B/T \leq 0.8$, and $B/T > 0.8$) for 
galaxies with $m_{F814W}(AB) \leq 26.0$.  The mean differences $d(B/T) 
\equiv (B/T)_{measured}-(B/T)_{input}$ in those three bins were 
$-$0.00, $-$0.04 and $-$0.14.  If we further divided our simulations 
into 0.1 $(B/T)$ bins, the mean differences going from $B/T = 0$ to 
$B/T = 1$ were 0.003 ($\sigma$=0.05), $-$0.03 (0.07), $-$0.04 (0.09), 
$-$0.05 (0.10), $-$0.02 (0.17), $-$0.02 (0.16), $-$0.04 (0.13), 
$-$0.04 (0.16), $-$0.08 (0.13), and $-$0.19 (0.24) with an average 
number of objects per bin of 45.  
The recovery success is lowest for $r_{hl}$ below about 0.2
which represents only 25\% of the galaxies with $m_{F814W}(AB) \leq 26.0$.
The larger deviation in the $B/T > 0.9$ bin has two main causes.  
First, we fitted a two-component model 
to all the objects.  Our model will therefore converge to a pure bulge 
model only when the signal-to-noise ratio is high enough to definitely 
rule out the presence of a disk component.  At lower signal-to-noise 
ratios, a disk component may be part of the model by being very small 
or very large.  Since we did not impose any correlation between 
structural parameters, both extremes were present in our simulations 
even though they may not be in real galaxies.  Second, the model can 
try to compensate for sky subtraction errors with a very low-surface 
brightness disk which would lead to an artificial decrease of the 
measured bulge fraction.  Finally, the larger deviation in the 
uppermost $B/T$ bin is also due to a smaller extent to the fact that 
we did not allow the $B/T$ to exceed 1.0 in our fits.  More objects 
will therefore be scattered out of the bin than into it.

As discussed below in Section~\ref{profano}, we found a number of 
galaxies in the HDF which could not be fitted by our model because 
they had surface brightness {\it flatter} than a pure exponential disk 
profile.  These objects were selected based on the following two 
criteria: (1) $B/T < 0.01$, and (2) the size of the 99\% $B/T$ 
confidence interval had to be less than 0.01.  We fitted objects 
satisfying both criteria using a pure S\'ersic profile given by 
equation~\ref{sersic}.  We hereafter refer to these objects as 
``S\'ersic'' galaxies.  Criterion (1) selected objects with the 
flattest surface brightness profiles allowed by our bulge/disk model, 
and criterion (2) measured the ``stubbornness'' of the model in its 
attempts to converge towards a profile flatter than a pure exponential 
disk.  A selection based solely on criterion (1) would probably have 
included pure disk exponential galaxies in our sample of S\'ersic 
galaxies.  The distribution of S\'ersic index values presented in 
Section~\ref{profano} shows that none of the S\'ersic galaxies we 
analyzed had $n = 1$.  The combination of both criteria successfully 
eliminated pure disk exponential galaxies.  However, criterion (2) was 
not ideal because the size of the confidence interval also depends on 
signal-to-noise: S\'ersic galaxies with low S/N ratios will have 
larger confidence intervals even if they exhibited the same 
``stubbornness'' to go beyond $n = 1$ as brighter S\'ersic galaxies.  
So, we do expect to be missing many faint S\'ersic galaxies.

Our S\'ersic selection criteria were tested by creating 300 simulated 
galaxies with pure S\'ersic profiles and structural parameters in the 
following ranges: $24.0 \leq m_{F814W}(AB) \leq 29.0$, $r_{e} \leq 0\arcsecpoint 
4$, $e \leq 0.7$ and $0.2 \leq n \leq 4.0$ (recall that $n = 4$ is a 
de Vaucouleurs profile, and $n = 1$ is the classical exponential disk 
profile).  These simulated galaxies were fitted with the combination 
of a de Vaucouleurs profile and a simple exponential profile.  The 
bottom part of Figure~\ref{sersic-simul-results} shows the measured 
$B/T$ versus input $n$ for simulations with $24.0 \leq m_{F814W}(AB) 
\leq 26.0$.  For $m_{F814W}(AB) \leq 26.0$, 41 simulations had $n < 
1.0$, but only 19 simulations passed our S\'ersic selection criteria.  
We therefore probably identified only half of the S\'ersic galaxies 
down to that magnitude limit in the HDF (see Section~\ref{profano} for further 
discussion).  For $m_{F814W}(AB) \leq 24.7$, the completeness fraction 
was 0.70.  None of the galaxies with input $n > 1$ met our S\'ersic 
selection criteria.

Our study of S\'ersic galaxies depends on our effectiveness to 
recognize them and on our ability to measure their S\'ersic index $n$ 
when $n \leq 4$.  We tested our ability to successfully recover 
S\'ersic index values by fitting pure S\'ersic models to the same set 
of S\'ersic galaxy simulations as above.  The top part of 
Figure~\ref{sersic-simul-results} shows the difference $dn$ between 
the observed S\'ersic index values and the input ones as a function of 
$n$.  For $0.0 \leq n_{input} \leq 1.0$, the mean value of $dn$ is 
$-$0.01 with an RMS of 0.04, and for $3.0 \leq n_{input} \leq 4.0$, 
the mean value of $dn$ is $-$0.37 with an RMS of 0.23.  These results 
clearly establish that we are able to measure changes in the structure 
of S\'ersic galaxies, and that galaxies with input $n > 1$ cannot be 
mistaken for S\'ersic galaxies.

\section{Comparison between Photometric Decomposition and Visual 
Classification} \label{comparison}

The HDF provides a unique database of galaxy images to study the 
morphological properties of a large sample of field galaxies.  VDB96 
has published a catalog of visually determined morphological 
classification for 19\% of the galaxies in the field which can be
compared to our quantitative classification.  Such a comparison is 
important as it provides a direct link between the population of 
nearby galaxies and high redshift galaxies, and this link is essential 
to our understanding of the evolution of field galaxies.  
The classification done by VDB96 was based on the DDO system, a system 
defined through the young-star richness of the disk, the presence of a bar, 
the central concentration of light and the quality and length of the 
arms of the galaxy.  A numerical system (Abraham \etal 1996b) 
accompanied the visual classification and was presented in VDB96.  
Objects such as peculiar galaxies or probable mergers were designated 
in the numerical system by the index vdB=7 or 8.  The rest of the 
numerical classification goes as follows: E/star:$-$1, E:0, 
E/S0/Sa:1, S0/Sa:2, Sa/Sab:3, Sb/S/Ir:4, Sc:5, and Ir:6.

VDB96 visually classified 271 galaxies in the HDF and derived the 
fractions of the different morphological types to be 30\% ellipticals
or lenticulars, 31\% spirals or irregulars, and 39\% unclassified
galaxies.  This fraction of HDF galaxies visually classified by VDB96  
is displayed in Figure~\ref{vdBfig} along with our 
$B/T$ parameter determination for comparison.  Each galaxy is 
represented by a circle, and this circle scales with the measured 
half-light radius of the galaxy.  If visual classification and our 
classification were in complete agreement, the points classified by 
VDB96 as elliptical galaxies would appear only in the upper left part 
of the diagram whereas the spirals should occupy only the lower right 
section.  A significant fraction, i.e.  14\%, of visually classified
elliptical galaxies are systems with a $B/T<0.5$.  It is interesting 
to note that all the galaxies classified by VDB96 as lenticular 
galaxies have $B/T<0.5$.  
The radially averaged profiles of some of the galaxies for which the 
VDB96 classification and ours differ are shown in 
Figure~\ref{profexamples}.  As the profiles suggest, these galaxies 
are disk-dominated galaxies but classified by VDB96 as ellipticals.
This difference in classification occurs for 40\% of small
round galaxies with half-light radii $<$ 0\arcsecpoint 31.  
The simulations indicate that GIM2D classifies these galaxies 
accurately (see Section~\ref{parrecov}).  These small objects 
account partly for the fact that we obtain a smaller number of early morphological types 
than the one determined using visual morphological classification methods.  

The VDB96 sample of galaxies has been classified by ABR96 using a 
quantitative classification system based on measurements of the 
central concentration and asymmetry of the galaxian light.  The $C-A$ 
classification developed by ABR96 is based on the results of Doi, 
Fukugita, \& Okamura (1993) describing how central concentration and 
mean surface brightness can be used to distinguish between 
early and late Hubble types.  The parameter $C$ is the ratio of the 
flux between the inner and outer isophotes of normalized empirically 
defined radii measured from the intensity-weighted second-order galaxy 
image moments, and the parameter $A$ is an asymmetry index measured by 
rotating and self-subtracting the galaxy image from itself.  A plot of 
the $C$ and $A$ indices calculated by ABR96 is displayed in 
Figure~\ref{abrahamfig}.  Each point is represented by a circle size 
proportional to the $B/T$ value from GIM2D. For complete agreement, 
all the circles with $B/T>0.5$ should lie in the region of the diagram 
defined as E/S0 and the circles with $B/T<0.5$ should belong to the 
central region of the figure labeled SPIRAL. Although all the 
bulge-dominated galaxies except for four fall in the part of 
the diagram expected, some members of the disk-dominated population are 
classified as elliptical galaxies.  Some, but not all, of these 
$B/T<0.5$ objects are the same objects identified by VDB96 as 
ellipticals.

There are four cases where bulge-dominated galaxies were classified as 
irregular/peculiar/merger (Irr/Pec/Mrg) in the $C-A$ classification.  
These galaxies are in the upper left corner of 
Figure~\ref{abrahamfig}.  The galaxies hd2$_-$0982$_-$1454 and 
hd4$_-$1589$_-$1175 have a companion which was identified in our 
sample as a separate object but where both objects were defined as one 
in the $C-A$ classification, making them compact Irr/Pec/Mrg objects.  
They were both classified by VDB96 with a classification index vdB=8.  
The galaxy hd4$_-$1075$_-$1749 is part of a quadruplet.  The grouping 
explains again the $C-A$ classification.  The VDB96 classification is 
vdB=0, consistent with a bulge-dominated galaxy.  The galaxy 
hd4$_-$0281$_-$1323 has irregular features and therefore is classified 
by ABR96 as Irr/Pec/Mrg but classified by VDB96 as vdB=0, consistent 
with our classification of a bulge-dominated galaxy.  It seems from 
this comparison that the VDB96 and ABR96 classifications do not 
consistently agree although they do agree in the instance that they 
are both classifying more galaxies as ellipticals than we find from 
our bulge/disk decomposition analysis.   

\section{Profile Anomalies in HDF Galaxies} \label{profano}

The surface brightness profiles of most galaxies we have analyzed with 
GIM2D could be modeled with the classical combination of a de 
Vaucouleurs profile and an exponential profile.  However, this model 
was insufficient to properly describe the photometric structure of a 
subsample of 82 S\'ersic galaxies that were selected based on the 
criteria described in Section~\ref{parrecov}.  The models of these galaxies 
converged to a pure exponential profile ($B/T=0$) with very small 99\%
confidence intervals.  This behavior of the model indicated that the 
profiles of those ``anomalous'' galaxies were flatter than any 
possible realization of the classical model.  These galaxies had 
profiles flatter than a pure exponential profile (see 
Figure~\ref{colgradexamples}) and could be described by a single-component 
S\'ersic model with index $0.2 \leq n \leq 1$.

The observed profile anomalies can be real intrinsic structural 
anomalies or they can be caused by, say, a vigorously star-forming 
population with a distribution very different from the underlying 
galaxy profile.  Although small perturbations on a smooth profile 
do not on average change the underlying profile parameters 
(see Section~\ref{asympardist}), large perturbations caused by merging, 
for example, can alter significantly the overall 2D profile shape.  
A blueing of the light from a pure disk galaxy in its 
central region can give rise to the kind of profile flattening 
observed in our subsample (see Figure~\ref{colgradexamples}) and yet 
have nothing to do with changes in the intrinsic structure of the 
galaxies.  It was therefore important to determine which anomalous 
galaxies had color gradients and which ones did not.  We created $V-I$ 
color images of the anomalous galaxies by dividing the F606W HDF 
images by the corresponding F814W images.  We derived $V-I$ profiles 
by azimuthally averaging their surface brightness along isophotal 
ellipses with axial ratios calculated by SExtractor.  Out of the 82 
galaxies, 34 galaxies showed little or no color gradient ($\delta 
(V-I)<0.2$).  The flat profiles of these galaxies thus represents real 
intrinsic structural anomalies.  Figure~\ref{sersic-n-nocolgrad} shows 
the distribution of S\'ersic index values $n$ for anomalous galaxies 
with no color gradient.  The distribution has a median $n$ value of 
0.62 and a dispersion of 0.18.  The S\'ersic galaxies exhibit an 
interesting diversity of visual morphologies.  Many are either mergers 
in progress or merger remnants as evidenced by tidal features such as 
tails.  The amount of merging appears to vary significantly.  Some 
mergers are between galaxies of roughly equal luminosities.  Others 
look like ``lit up Christmas trees'': many small, bright knots appear 
to be falling the potential well of a larger, central component.  
Figure~\ref{mosaic_sersic} shows a mosaic of four representative 
S\'ersic galaxy $F814W$ images.

\section{Discussion} \label{discuss}

We have fitted the surface brightness profiles of 1639 objects in the 
HDF and have obtained reliable quantitative morphological 
classification of 522 objects down to $m_{F814W}(AB) = 26.0$.  The 
GIM2D classification has been tested through simulations and compared 
to the visual classification method of VDB96 and ABR96.  Five effects 
were modeled: the galaxy selection function of the SExtractor 
detection algorithm, the parameter recovery success of GIM2D, 
the effect of asymmetric structures on the measured parameters, the 
selection criterion for galaxies with S\'ersic profiles, and the 
recovery success of the S\'ersic index for these galaxies.  We found 
that incompleteness was not a major problem down to the above 
magnitude limit.  However, the galaxy selection function showed a 
clear bias against pure disk systems whereas all bulge-dominated 
galaxies appeared to have been found.  This bias is due to the disk 
profile being less concentrated than the bulge profile.  The following 
results have emerged from the simulations and profile analysis of HDF 
galaxies.

The dominant contribution to the galaxy population in the HDF comes 
from the disk-dominated galaxies.  There are no large ($r_{hl} \ge 
0\arcsecpoint 7$) bulge-dominated systems in our sample of the HDF, 
and the fraction of bulge-dominated galaxies ($B/T > 0.5$) down to 
$m_{F814W}(AB) = 26.0$ is 8\%.  This is a much smaller percentage than 
found by VDB96 and ABR96 with visual classification.  The discrepancy 
is even more pronounced if we associate VDB96's E/S0 systems with $B/T 
> 0.8$.  We find that the discrepancy between our low percentage of 
early-type galaxies and the larger percentage found by visual 
classification is due to the difference in classification of 40\% of 
small round galaxies with half-light radii $<$ 0\arcsecpoint 31.  
Although these objects are visually classified as elliptical galaxies, 
we find that they are disk-dominated with bulge fractions $<$ 0.5.
The simulations indicate that GIM2D classifies these galaxies accurately.  

This result emphasizes two obvious problems of visual classification.  
First, the visual method is neither reliable nor reproduceable.  In general, the 
subjectivity of visual classification makes it impossible to measure 
or even simply determine in a consistent way the systematic errors 
associated with this type of classification.  Second, our quantitative system
of classification is based on a set of measurements that only roughly 
maps onto the DDO system due to the fact that $B/T$ is
an important component of that system.  Our different classification 
results on the HDF suggest that profile shape has only a very indirect 
relationship with visual morphological classification.  It is not possible
to map galaxies directly onto the DDO system by measuring profile
parameters.   Therefore, it is difficult to compare samples of 
distant and nearby galaxies, which have predominantly been classified visually.

Assuming a one-to-one correspondence between visual and bulge/disk 
classification for the nearby sample of galaxies, we compare our 
revised number for the frequency of morphological types in the HDF 
with the local sample as quoted in VDB96.  Since the HDF galaxy 
redshift distribution peaks at $z\sim0.5$ (Cohen \etal 1996), we are 
measuring a {\it decrease} in the number of bulge-dominated galaxies 
as a function of look-back time.  This agrees with the absence of 
passively evolving ellipticals in deep optical and near-infrared 
surveys (Zepf 1997).  Up to a redshift of $z\sim1$, stellar synthesis 
models (Bruzual \& Charlot 1993) predict that the elliptical galaxies 
should stay red in $V-I$ color and therefore should not be 
disappearing from our sample at these low redshifts, unless they form 
in environments containing a significant amount of dust.  If real, 
this decrease may be due to merging and at least some of the elliptical galaxies 
will be formed by the coalescence of colliding galaxies of different 
morphological types.  It is also possible that spheroids form by the collapse 
of many subclumps (Kauffmann \etal 1993; Navarro \etal 1995; Cole 
\etal 1994) and that these subclumps and substructure have different 
morphologies.  By collecting color as well as morphological 
information for the HDF galaxies, it is possible to determine if the 
morphological evolution is also responsible for the excess number of 
blue galaxies at faint magnitudes in redshift surveys or if a single 
population of galaxies is evolving in color.  This study, including 
the use of redshift information available now on the HDF galaxies, 
will make up the content of a future paper.

With redshift information, we can study the morphological evolution of 
galaxies and test the claim that high redshift galaxies are in general 
very compact, with $r^{1/4}$ law radial profiles, and with scales 
comparable to the cores of present-day luminous galaxies (Giavalisco 
\etal 1996a; Giavalisco \etal 1996b; Steidel \etal 1996a; Steidel 
\etal 1996b).  Because of these properties, these galaxies are thought 
to be most likely the progenitors of bulges or normal elliptical 
galaxies.  The $z>3$ galaxies with measured spectroscopic redshifts of 
Lowenthal \etal (1997) have small sizes, with half-light radii in the 
range 1.5-3 $h_{50}^{-1}$ kpc (for $q_0=0.05$), but show a wide range 
of morphologies.  As Lowenthal \etal (1997) suggest, the small sizes 
of these objects and their morphology are also consistent with other 
scenarios of galaxy formation and evolution.  The substructure seen in 
some objects could be due to the hierarchical formation of the galaxy 
beginning with subclumps or they could be isolated knots of star 
formation.

If all the high redshift ($z>2$) objects are indeed bulge-dominated 
galaxies, it appears from our analysis that the morphological 
distribution of galaxies in the HDF does not agree with these 
findings.  In Figure~\ref{redshift_spec}, we present the bulge 
fraction of the sample of 61 galaxies with measured spectroscopic 
redshifts (Cohen \etal 1996, Lowenthal \etal 1997, Steidel \etal 
1996a, and Zepf \etal 1997).  The lack of objects between $z\sim1-2$ 
is due to the observing strategy and target selection of the different 
groups and the difficulty of spectral identification.  All the 
galaxies range in sizes from $r_{hl}=$0\arcsecpoint 10 to 
$r_{hl}=$1\arcsecpoint 12 and have redshifts from $z=0.129$ to 
$z=3.368$.  Assuming $q_0=0.5$, the 15 objects with $z>2.0$ have 
half-light radii in the range 0.73-2.99 $h_{50}^{-1}$ kpc and only one 
third of the objects are bulge-dominated galaxies.  Of these 
bulge-dominated galaxies, hd4$_-$1588$_-$1174, the largest, is at 
redshift $z=2.803$ with size $r_e=$4.19 $h_{50}^{-1}$ kpc, comparable 
to the size of a normal elliptical galaxy (Bender \etal 1992).  The 
smallest bulge at $z=3.21$, hd4$_-$1075$_-$1749, which is part of a 
quadruplet (see Section~\ref{comparison}), has $r_e=$1.63 
$h_{50}^{-1}$ kpc, corresponding to the size of a dwarf elliptical.  
These two galaxies have high luminosities $L>L^\ast$ (Lowenthal \etal 
1997).  The objects selected by Lowenthal \etal (1997) are at $z>2.0$ 
and have $B/T\leq0.54$.  All but two have $B/T<0.2$, so the majority 
of these galaxies have dominant exponential profiles and do not fit 
the description of cores of proto-spheroids unless they merge later on 
to form elliptical galaxies.  These objects are small compared with 
present day disks, with a range in their half-light radii of 0.73-2.60 
$h_{50}^{-1}$ kpc.

We have found 82 galaxies with profiles shallower than an exponential 
profile.  Fitting pure S\'ersic profiles to those objects and 
excluding those with large color gradient, we found that the S\'ersic 
indices of the remaining 34 galaxies had a median value of 0.62 and a 
dispersion of 0.18.  Many of these S\'ersic galaxies shows evidence of 
mergers. We intend to study these S\'ersic galaxies in greater details 
in a future paper.

The outcome of our analysis on the HDF images are summarized here: (1) 
Our morphological analysis of 522 galaxies in the field indicates that 
the spheroidal or bulge-dominated galaxies make up for only a small 
fraction, i.e. 8\%, of the galaxy population down to m$_{F814W}(AB)$ =
26.0.  (2) We showed that the large fraction of early-type systems in the
VDB96 and ABR96 sample is due to the difference in classification of
small round galaxies with half-light radii $<$ 0\arcsecpoint 31.  Although 
these objects are visually classified as elliptical galaxies, we find that 
they are disk-dominated with bulge fractions $<$ 0.5.
(3) We found a majority of disk-dominated galaxies in our high redshift 
($z>2$) sample and only a small fraction of spheroids.  (4) We 
observed galaxies with profiles flatter than a pure exponential 
profile and many of these objects show signs of merging.

\acknowledgments

F.R.M. would like to acknowledge support from the HST NASA grant 
\#AR-07523.01 provided by T.J. Broadhurst.  L.S. gratefully 
acknowledges financial support from the Natural Sciences and 
Engineering Research Council of Canada through a Postdoctoral 
Fellowship and support from HST grant \#AR 06337.08-94A by D.C Koo.  We 
would like to thank M. Bolte, J.R. Graham and A.C. Phillips for 
helpful discussions and comments.  We are also grateful to our referee 
R.G. Abraham for his comments and suggestions on improving the paper.

\clearpage 
\figcaption[marleau98_fig1.ps]{{\it Left}: The top left quarter of the WFPC2-chip 4 
F814W HDF image, displaying a field of view of 36\arcsecpoint 8 $\times$ 
36\arcsecpoint 0.  {\it Right}: The residual image created by GIM2D 
after detailed bulge/disk decompositions have been performed on all 
galaxies in the field.
\label{hd4image}}

\figcaption[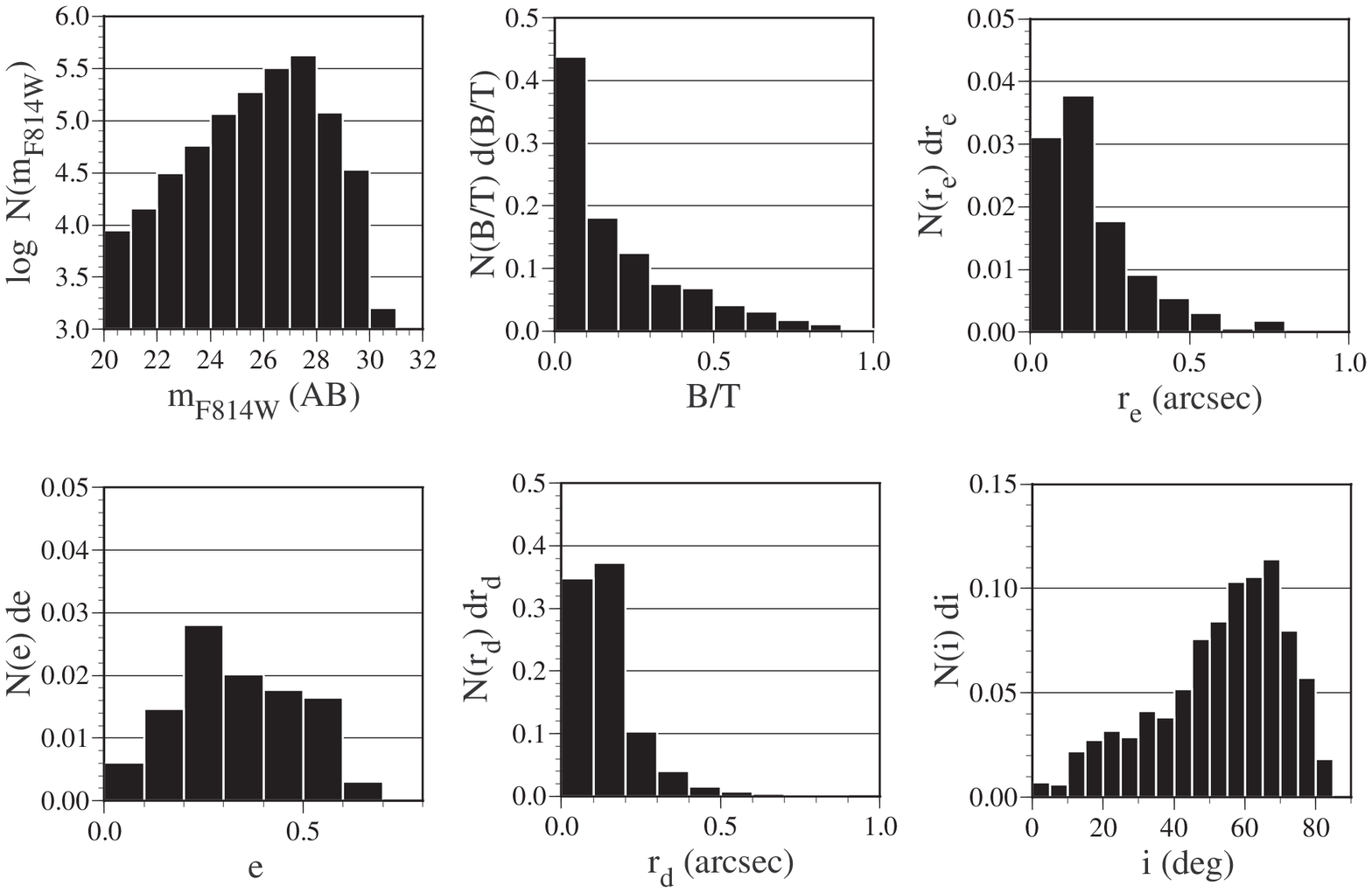]{Observed structural parameter 
distributions of the HDF galaxies for all 1639 objects analyzed with 
GIM2D. All distributions except $N(m_{F814W}(AB))$ have been 
normalized by the total number of objects in our catalog (1639  
objects).  $N(m_{F814W}(AB))$ is in mag$^{-1}$ deg$^{-2}$.
\label{observed-dist-all}}

\figcaption[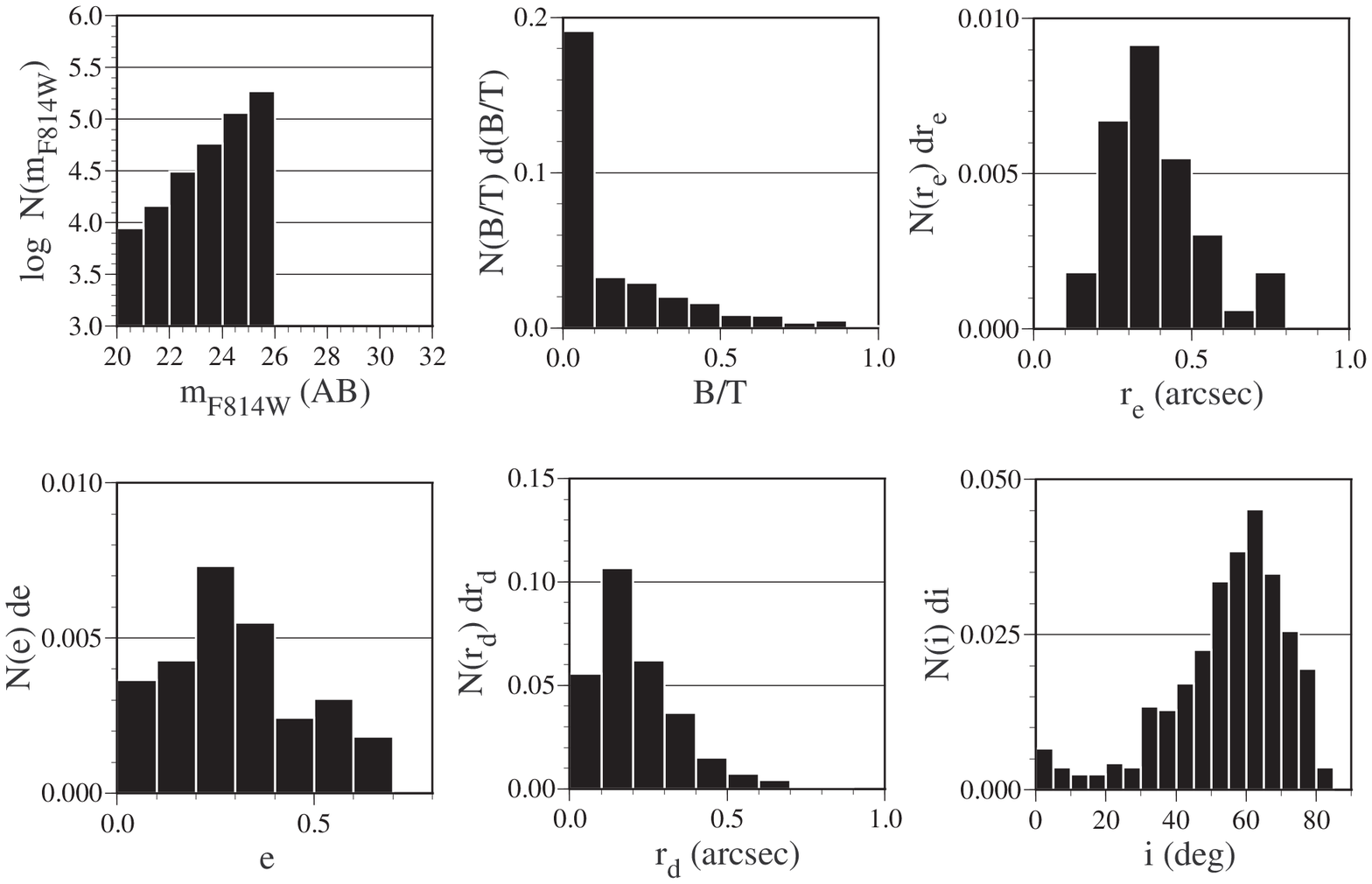]{Observed structural parameter 
distributions of the HDF galaxies with $m_{F814W}(AB) \leq 26.0$.  All 
distributions except $N(m_{F814W}(AB))$ have been normalized by the 
total number of objects in our catalog (1639 objects).  
$N(m_{F814W}(AB))$ is in mag$^{-1}$ deg$^{-2}$.
\label{observed-dist-tmlt26}}

\figcaption[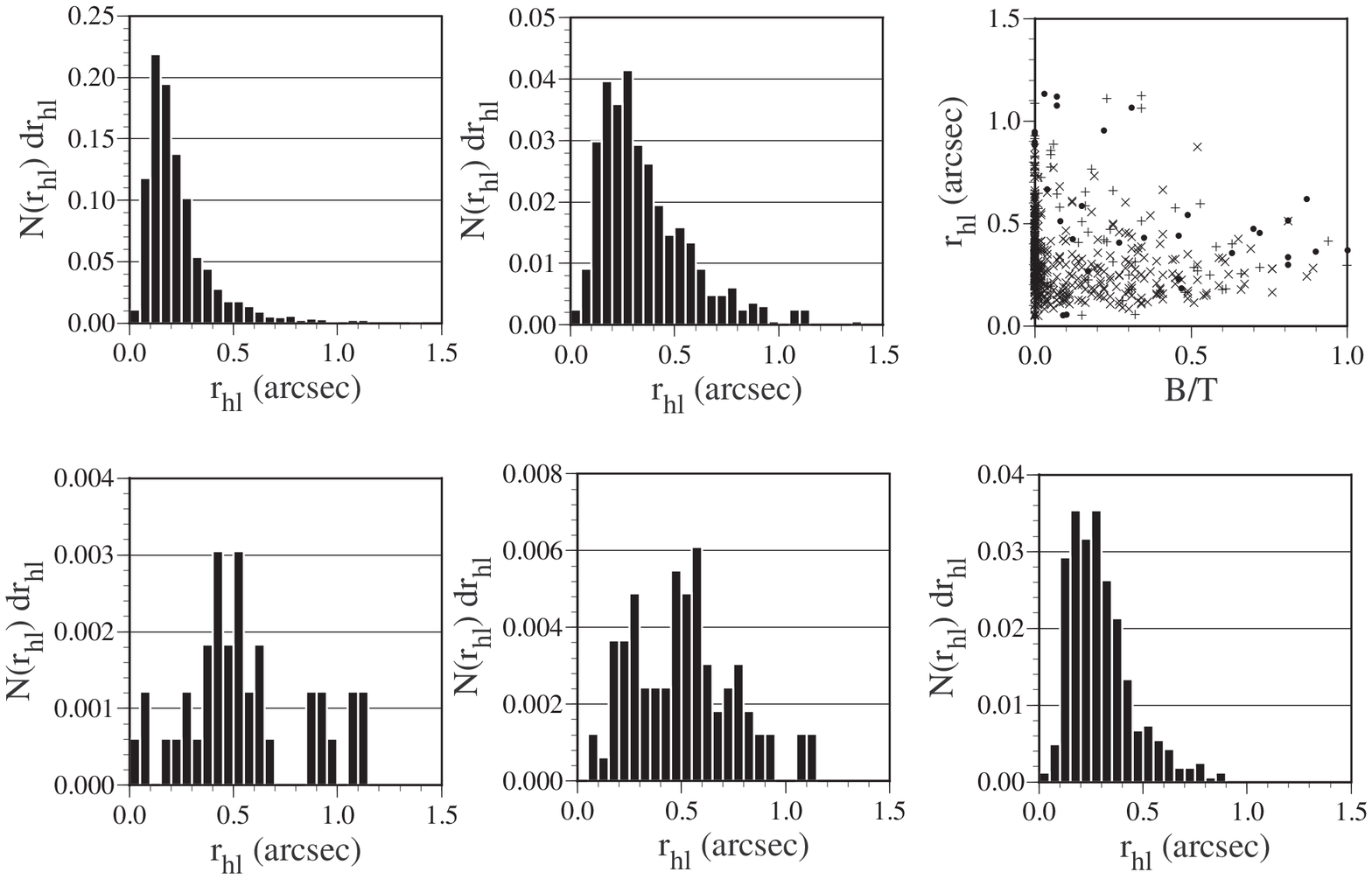]{Half-light radius distributions 
of the HDF galaxies for different $m_{F814W}(AB)$ cuts.  All 
distributions have been normalized by the total number of objects in 
our catalog (1639 objects).  Clockwise from top left-hand corner: 
$N(r_{hl})$ for all 1639 objects, $N(r_{hl})$ for $m_{F814W}(AB) \leq 
26.0$, $r_{hl}$ versus $B/T$ for $21.0 \leq m_{F814W}(AB) \leq 22.6$ 
(filled circles), $22.6 < m_{F814W}(AB) \leq 24.0$ (pluses), $24.0 < 
m_{F814W}(AB) \leq 26.0$ (crosses), $N(r_{hl})$ for $24.0 < 
m_{F814W}(AB) \leq 26.0$, $N(r_{hl})$ for $22.6 < m_{F814W}(AB) \leq 
24.0$, $N(r_{hl})$ for $21.0 < m_{F814W}(AB) \leq 22.6$.
\label{observed-dist-hlr}}

\figcaption[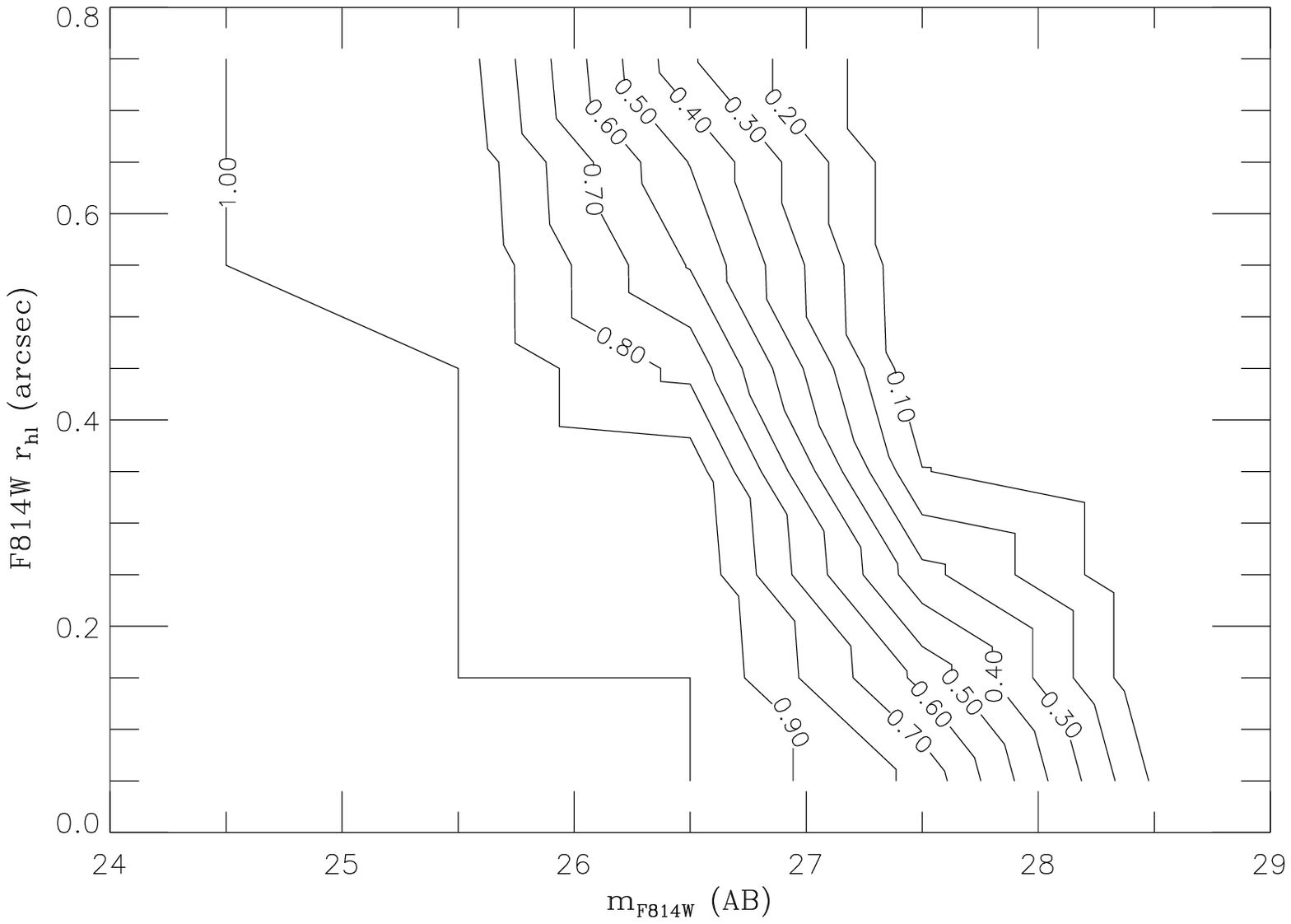]{Contour plot of the SEXtractor selection 
function $S(\omega)$ as a function of the F814W half-light radius 
$r_{hl}$ and the total F814W AB magnitude for the HDF. $S(\omega)$ was 
constructed with 66000 galaxy models spanning a wide range of 
structural parameters.  The detection threshold was 1.5$\sigma$, and 
the minimum object detection area was 30 pixels.
\label{selfplot}}

\figcaption[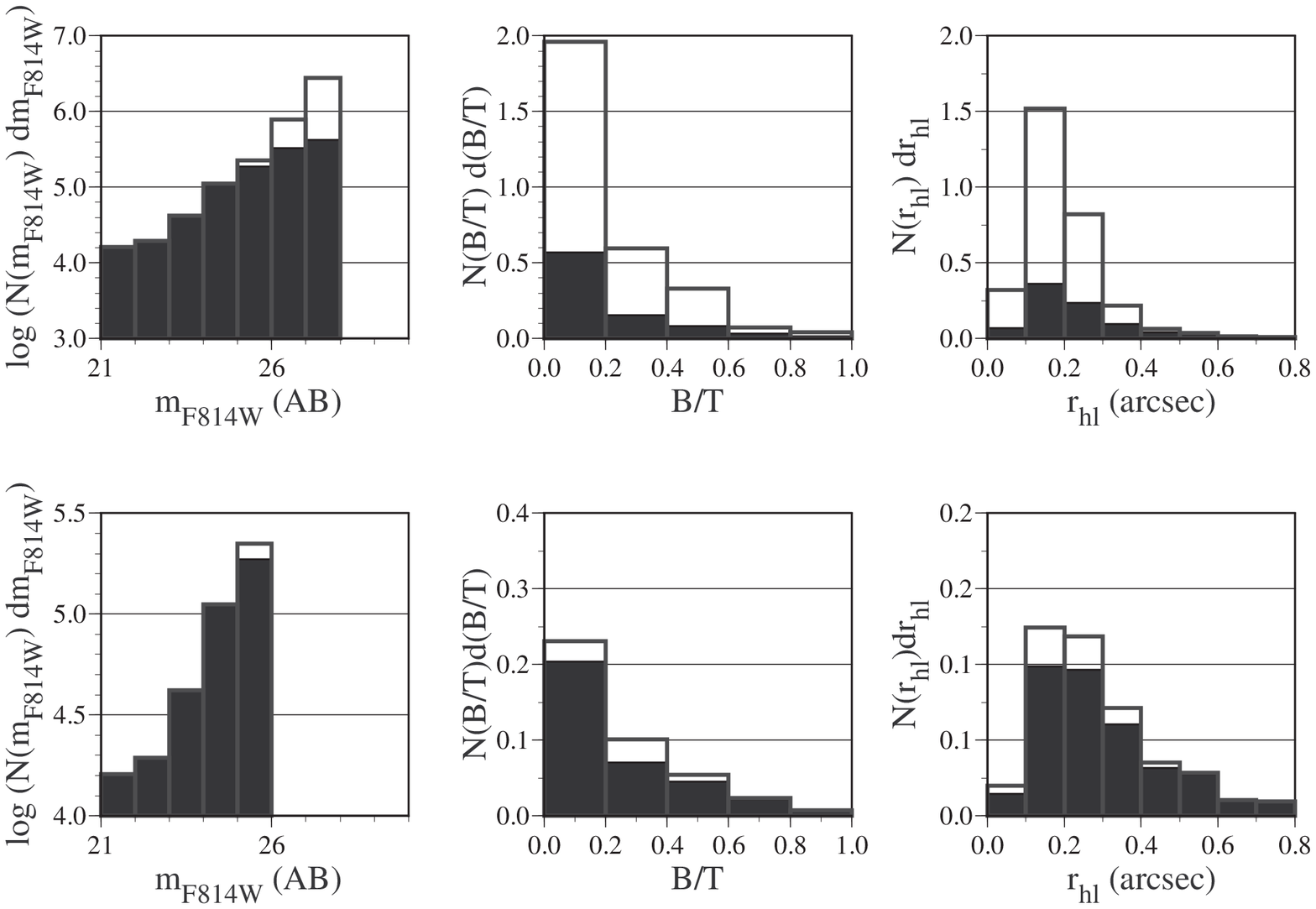]{Observed ({\it solid histogram}) 
and intrinsic ({\it dashed histogram}) structural parameter 
distributions for $m_{F814W} (AB)$, $B/T$, and $r_{hl}$ calculated 
over the range $21.0 < m_{F814W}(AB) \leq 29.0$ for $S_{lim} = 0.1$ 
({\it top}) and $S_{lim} = 0.5$ ({\it bottom}).  Notice that objects 
with magnitudes down to $m_{F814W}(AB) = 28.0$ are detected for the 
$S_{lim} = 0.1$ but for the most stringent limit of $S_{lim} = 0.5$, 
only objects in the brighter bins with $m_{F814W}(AB) \leq 26.0$ are
detected.  
\label{intrinsic-dist-2129}}

\figcaption[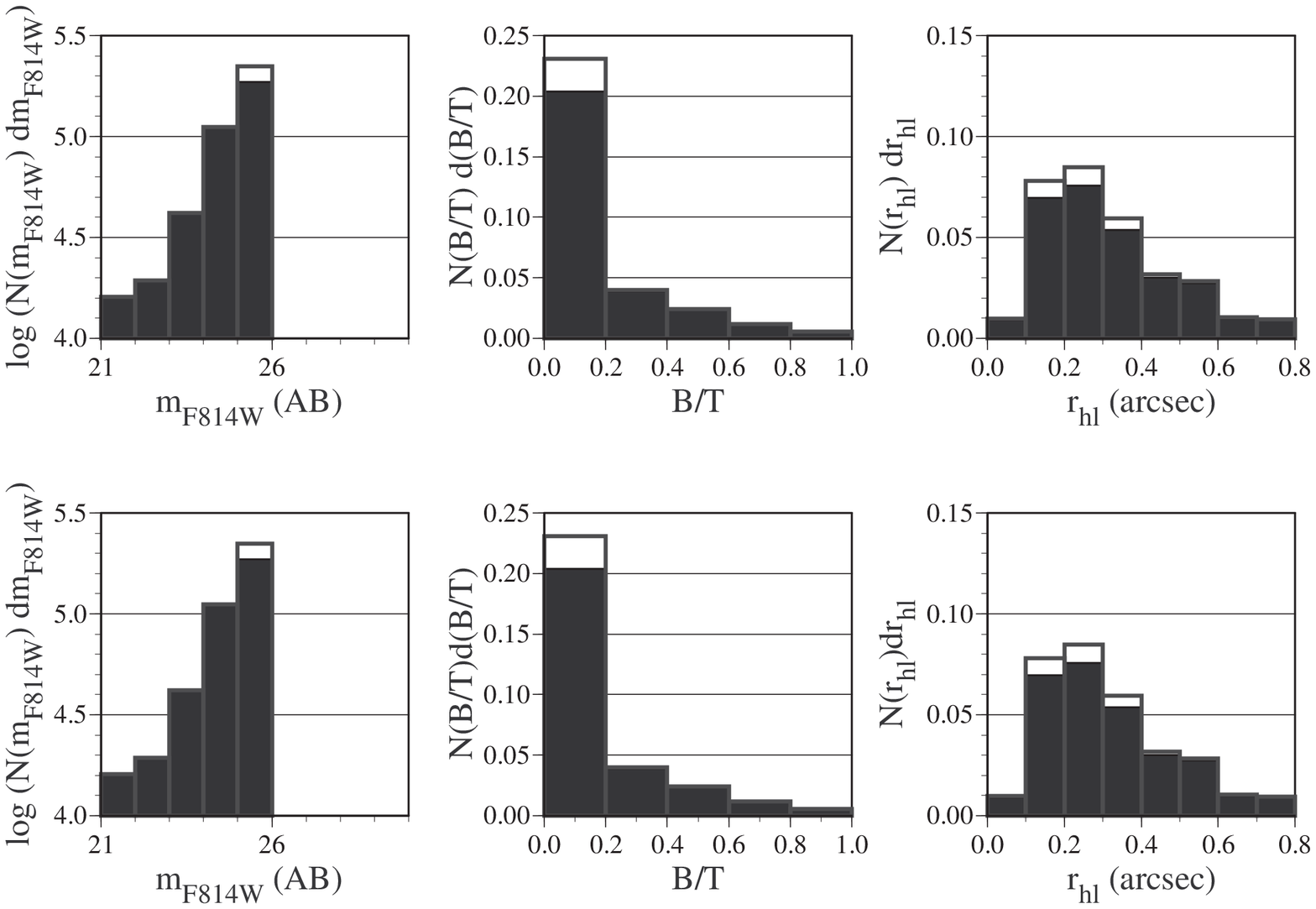]{Observed ({\it solid histogram}) and 
intrinsic ({\it dashed histogram}) structural parameter 
distributions for $m_{F814W} (AB)$, $B/T$, and $r_{hl}$ calculated 
over the range $21.0 < m_{F814W}(AB) \leq 26.0$ for $S_{lim} = 0.1$ 
({\it top}) and $S_{lim} = 0.5$ ({\it bottom}).  
\label{intrinsic-dist-2126}}

\figcaption[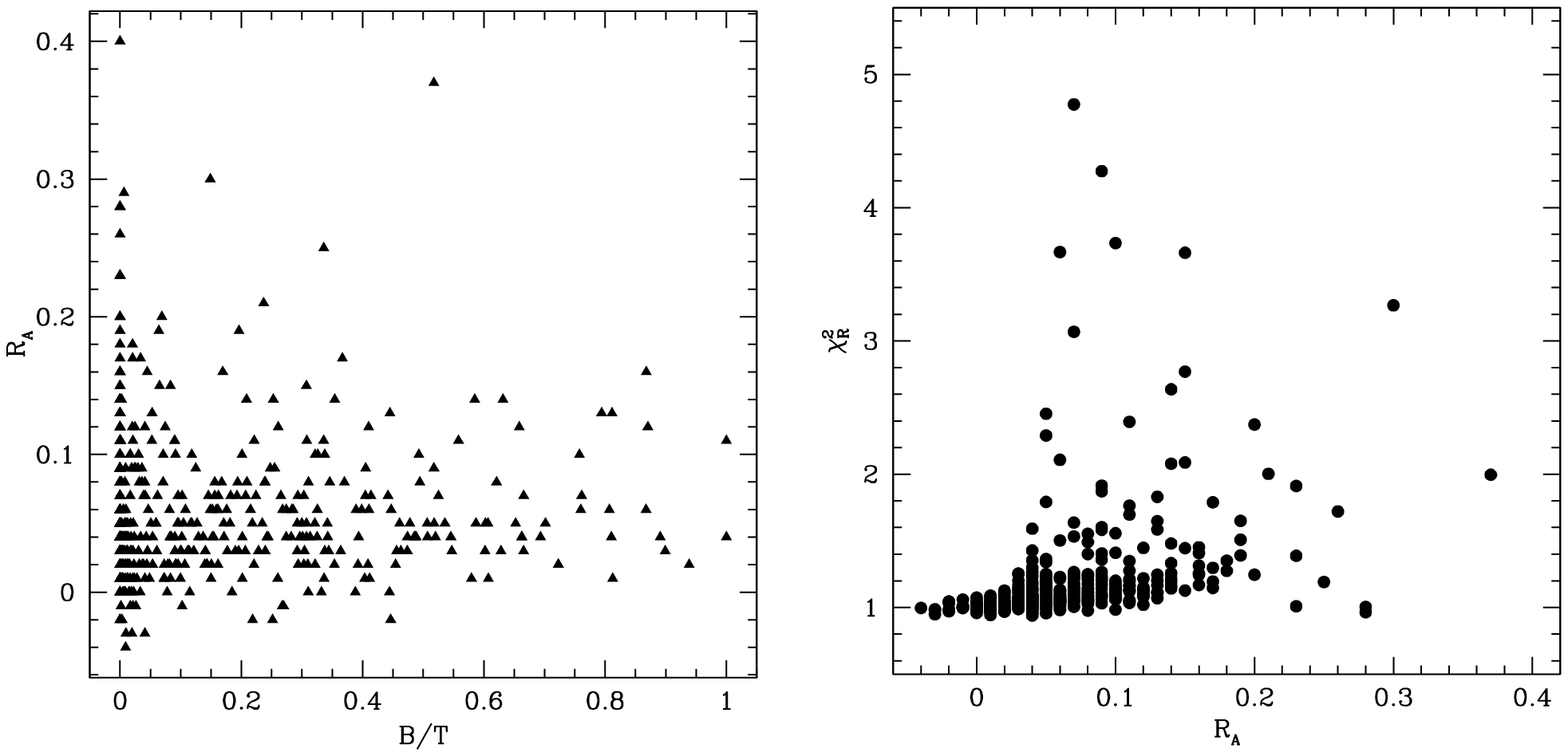]{
{\it Left}: Asymmetric residual fluxes index $R_A$ as a function of $B/T$ 
for our sample of 522 HDF galaxies with $m_{F814W}(AB) \leq 26.0$.  
$R_A$ is computed using an aperture of size 2$r_{hl}$ and remains smaller 
than $\sim$20\% for the majority of galaxies.  
{\it Right}: A measure of the residual fluxes $\chi^2_R$ as a function 
of the degree of asymmetry $R_A$ for the same sample of galaxies.   
\label{ra12}}

\figcaption[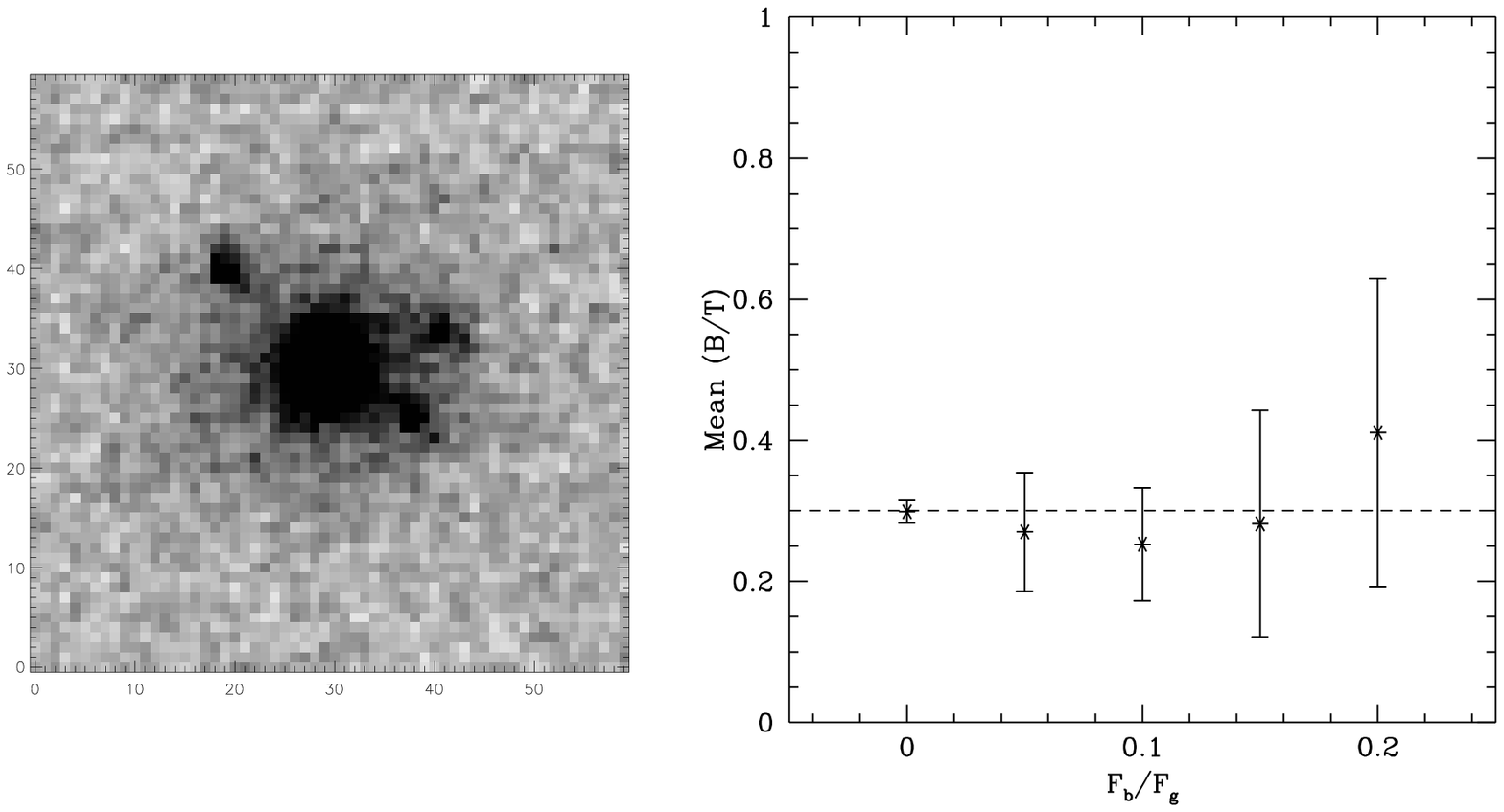]{ 
{\it Left}: Example of a 
2\arcsecpoint 4 $\times$ 2\arcsecpoint 4 simulated galaxy image 
with asymmetric features parametrized by $F_b=0.10$, $n_b=5$ and $r_b=1.5r_{hl}$.  
The simulations were generated for the smooth
component parameters $m_{F814W}(AB)=24.0$, $B/T=0.3$, $r_{e}=0\arcsecpoint 12$,
$e=0.2$, $r_{d}=0\arcsecpoint 32$, $i=20\deg$, $\phi_b=\phi_d=60\deg$,
and $n=4.0$.
{\it Right}: Mean measured $B/T$ with 1$\sigma$ error bars 
for the asymmetric parameters $n_b=5$, $r_b=1.5r_{hl}$ and 
the five discrete flux levels $F_b=$0.0, 0.05, 0.10, 0.15, 0.20.  
\label{simulb}}

\figcaption[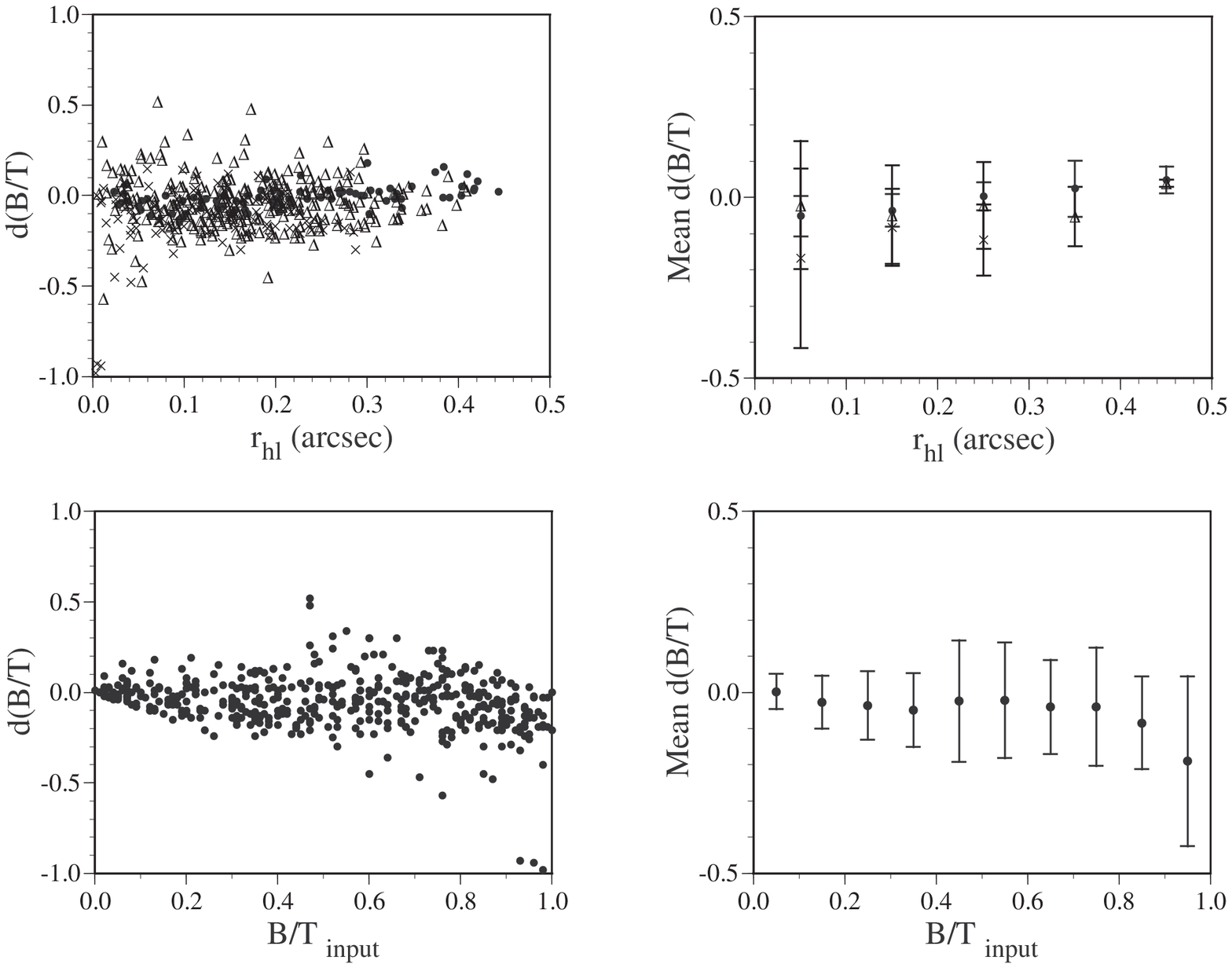]{{\it Top and left}: The 
difference $d(B/T)$ between measured and input bulge fractions versus 
seeing-deconvolved half-light radius for three different bulge 
fraction intervals (0.0 $< B/T \leq$ 0.2 (solid circles), 0.2 $< B/T 
\leq$ 0.8 (open triangles), and 0.8 $< B/T$ (crosses)). {\it Top and 
right:} Mean difference $d(B/T)$ between measured and input bulge 
fractions versus seeing-deconvolved half-light radius with 1$\sigma$ 
error bars.  {\it Bottom and left:} The difference $d(B/T)$ between 
measured and input bulge fractions versus the input bulge fraction.  
{\it Bottom and right:} Mean difference $d(B/T)$ between measured 
and input bulge fractions versus input bulge fraction with 1$\sigma$ 
error bars.  All simulations with $m_{F814W}(AB) \leq 26.0$ were 
included.
\label{simul-results-hlr+bt}}

\figcaption[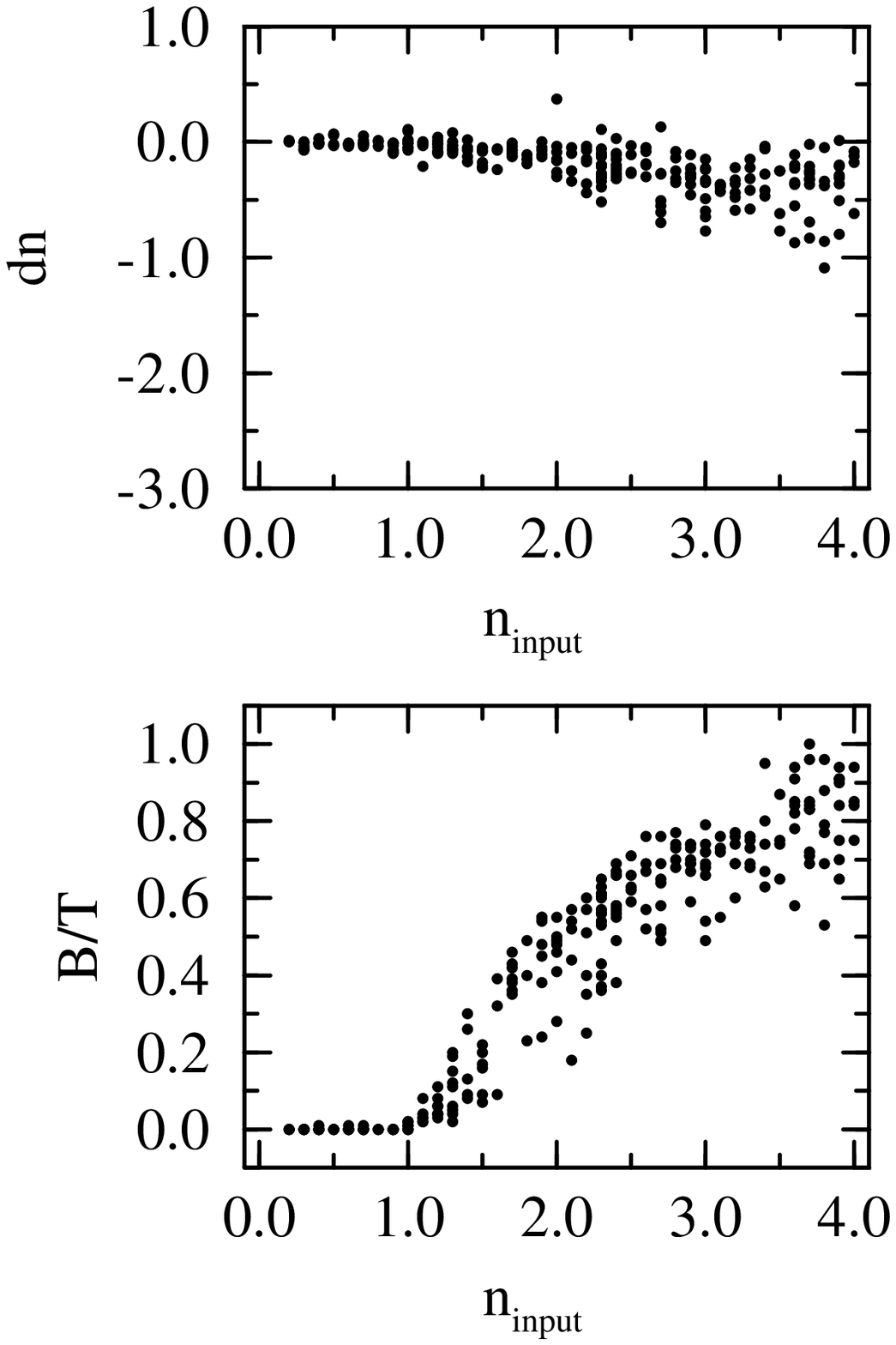]{{\it Top}: Difference $dn$ 
between measured and input S\'ersic indexes versus input S\'ersic 
index obtained by fitting pure S\'ersic models to 300 pure S\'ersic 
profile simulations.  {\it Bottom}: Measured bulge fraction $B/T$ versus 
input S\'ersic index $n$ obtained by fitting bulge/disk model to 300 
pure S\'ersic profile simulations.  All simulations with 
$m_{F814W}(AB) \leq 26.0$ were included.
\label{sersic-simul-results}}

\figcaption[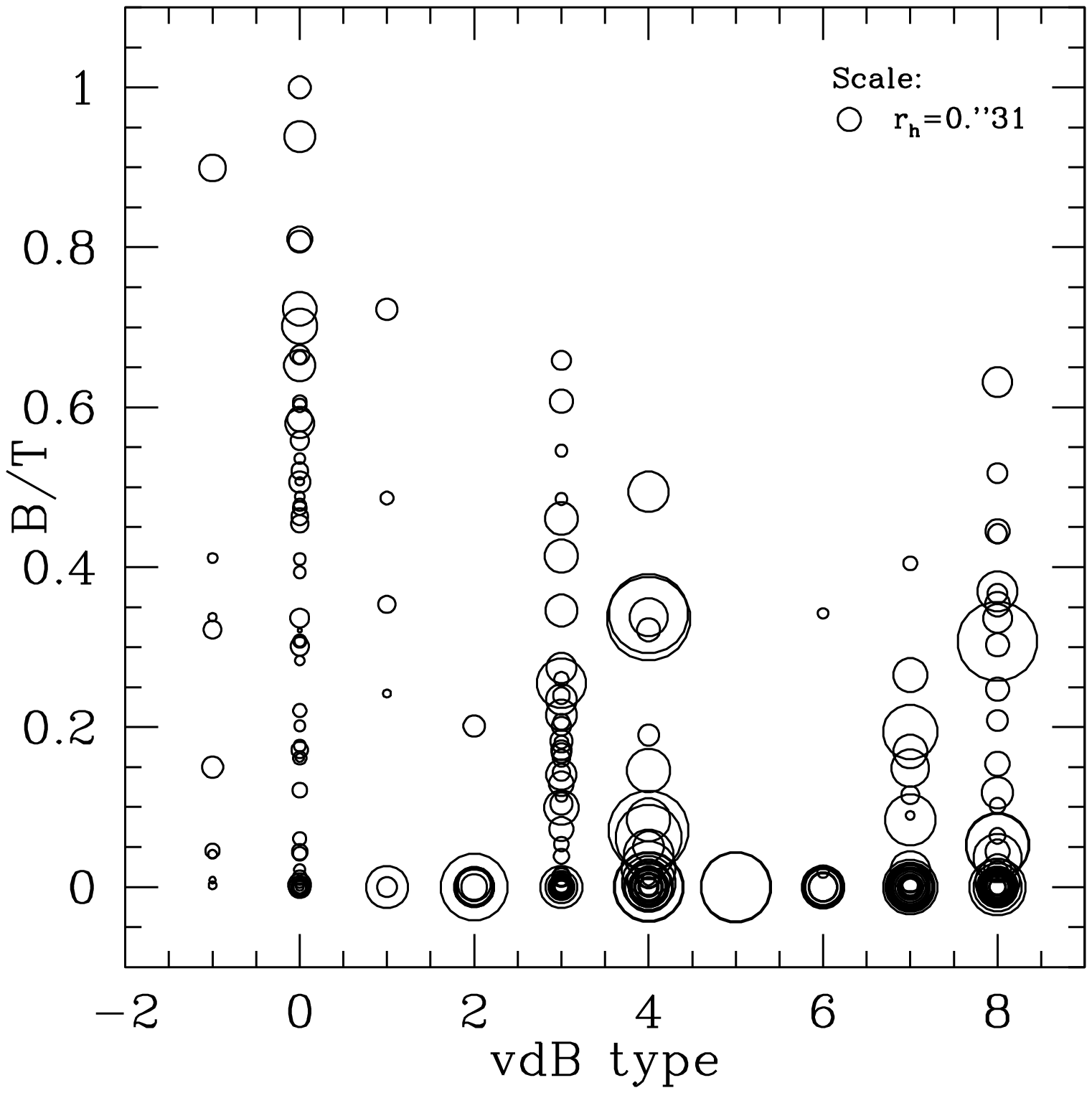]{Comparison of the parameter $B/T$ with the 
morphological type vdB derived from VDB96.  The size of the circles 
is proportional to $r_{hl}$.  If visual classification and our
classification were in complete agreement, the points classified by
VDB96 as elliptical galaxies would appear only in the upper left part
of the diagram whereas the spirals should occupy only the lower right
section.\label{vdBfig}}

\figcaption[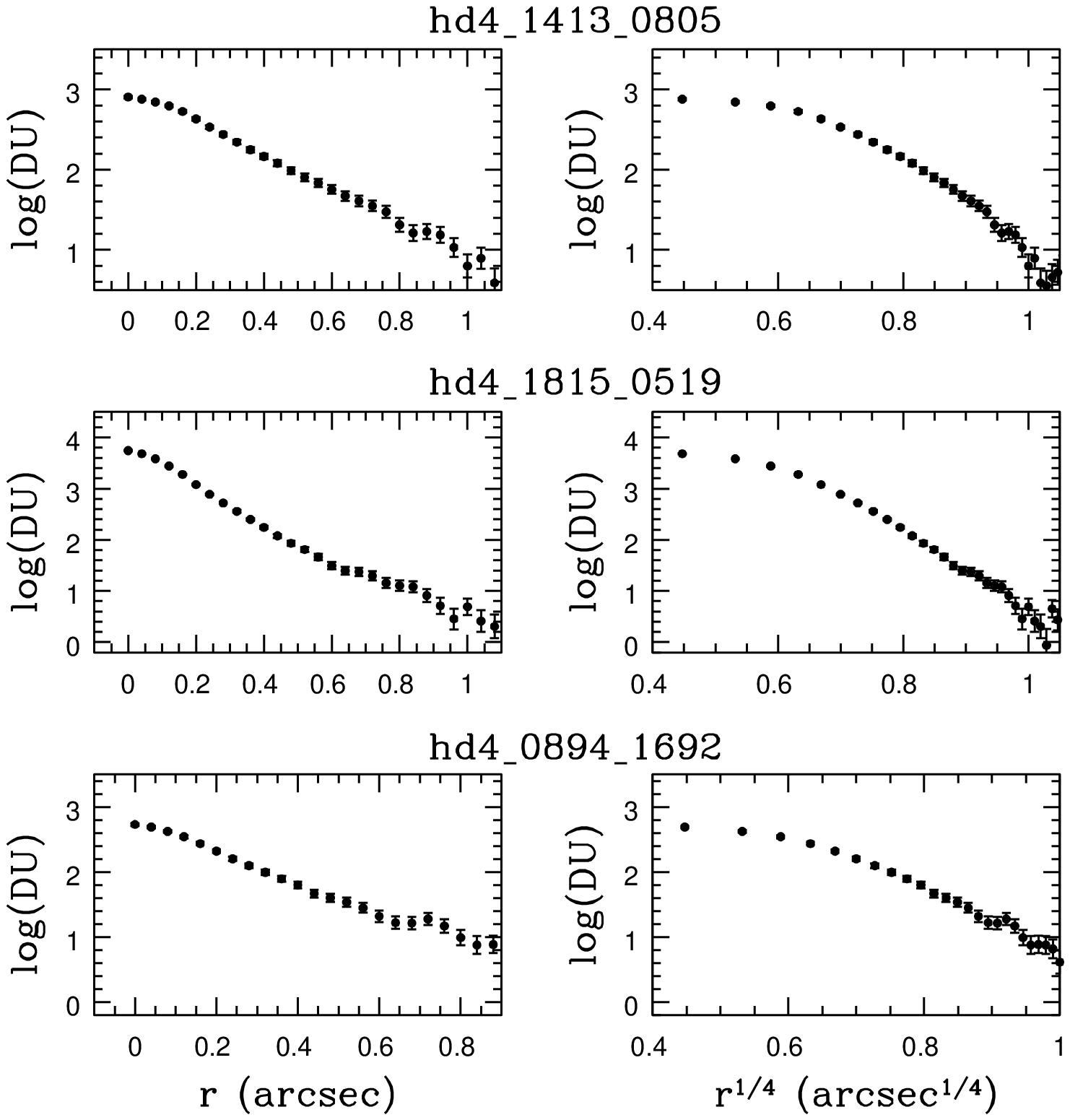]{Examples of the difference in classification of small 
round galaxies.  The three rows, from top to bottom, are the radially 
averaged surface brightness profiles of the galaxy 
hd4$_-$1413$_-$0805, hd4$_-$1815$_-$0519 and hd4$_-$0894$_-$1692.  The 
points are drawn with their 1$\sigma$ Poissonian error bars.  They are 
assigned a VDB96 classification type vdB=0 but a bulge/disk 
decomposition suggests increasing values of $B/T$ = 0.0029, 0.2833 and 
0.3365.  These galaxies are clearly dominated by an exponential 
profile component and the difference in classification appears related to the fact 
that these galaxies are small and round, with $r_{hl}$ = 0\arcsecpoint 
31, 0\arcsecpoint 12 and 0\arcsecpoint 25, respectively.
\label{profexamples}}

\figcaption[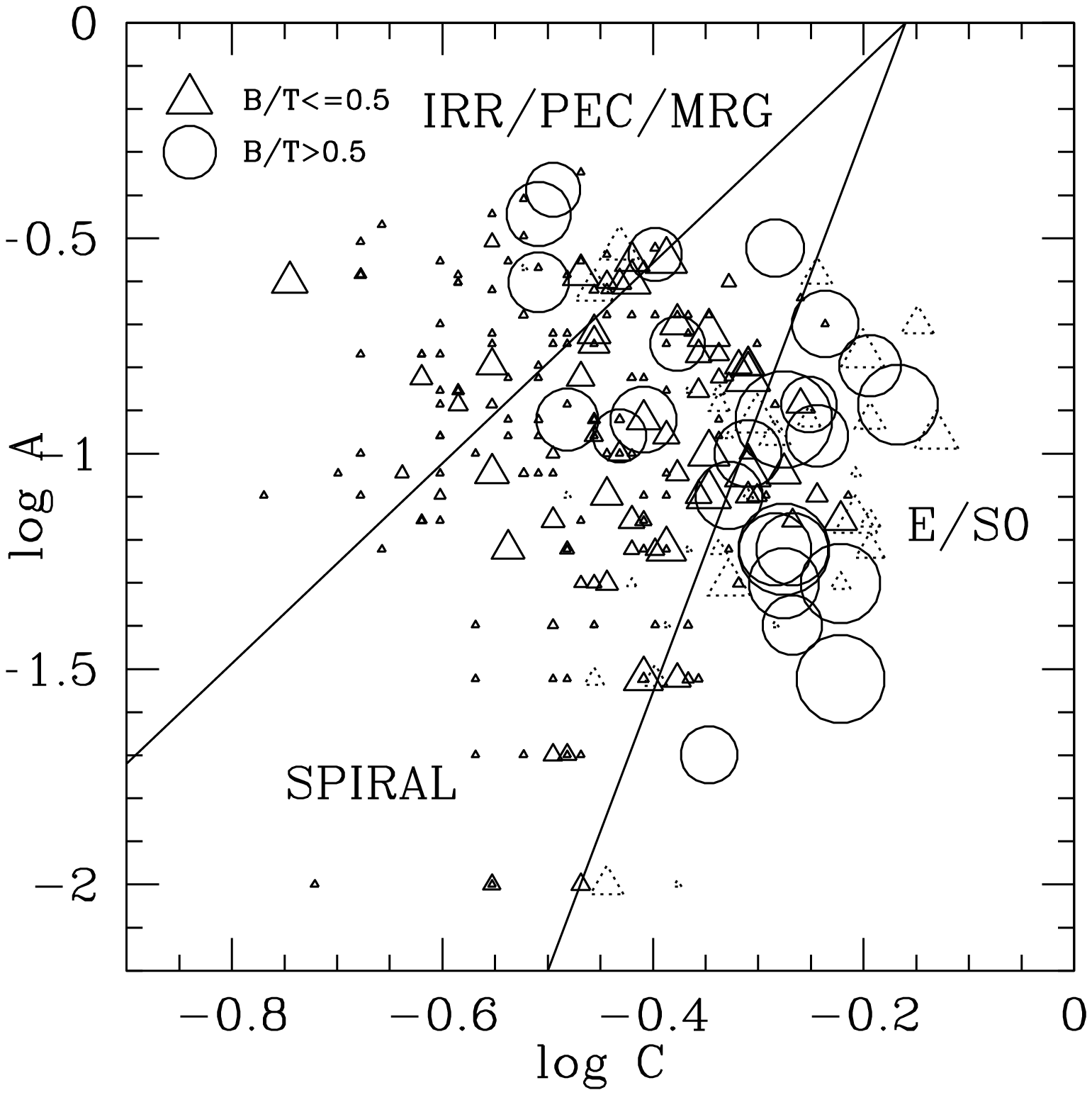]{Comparison with the $C-A$ classification of ABR96.  
The triangles and circles represent objects with $B/T \leq 0.5$ and $B/T > 0.5$, respectively,
with sizes proportional to $B/T$.  The dotted triangles 
are for galaxies with $B/T<0.5$ but classified by VDB96 as vdB=0.  \label{abrahamfig}} 

\figcaption[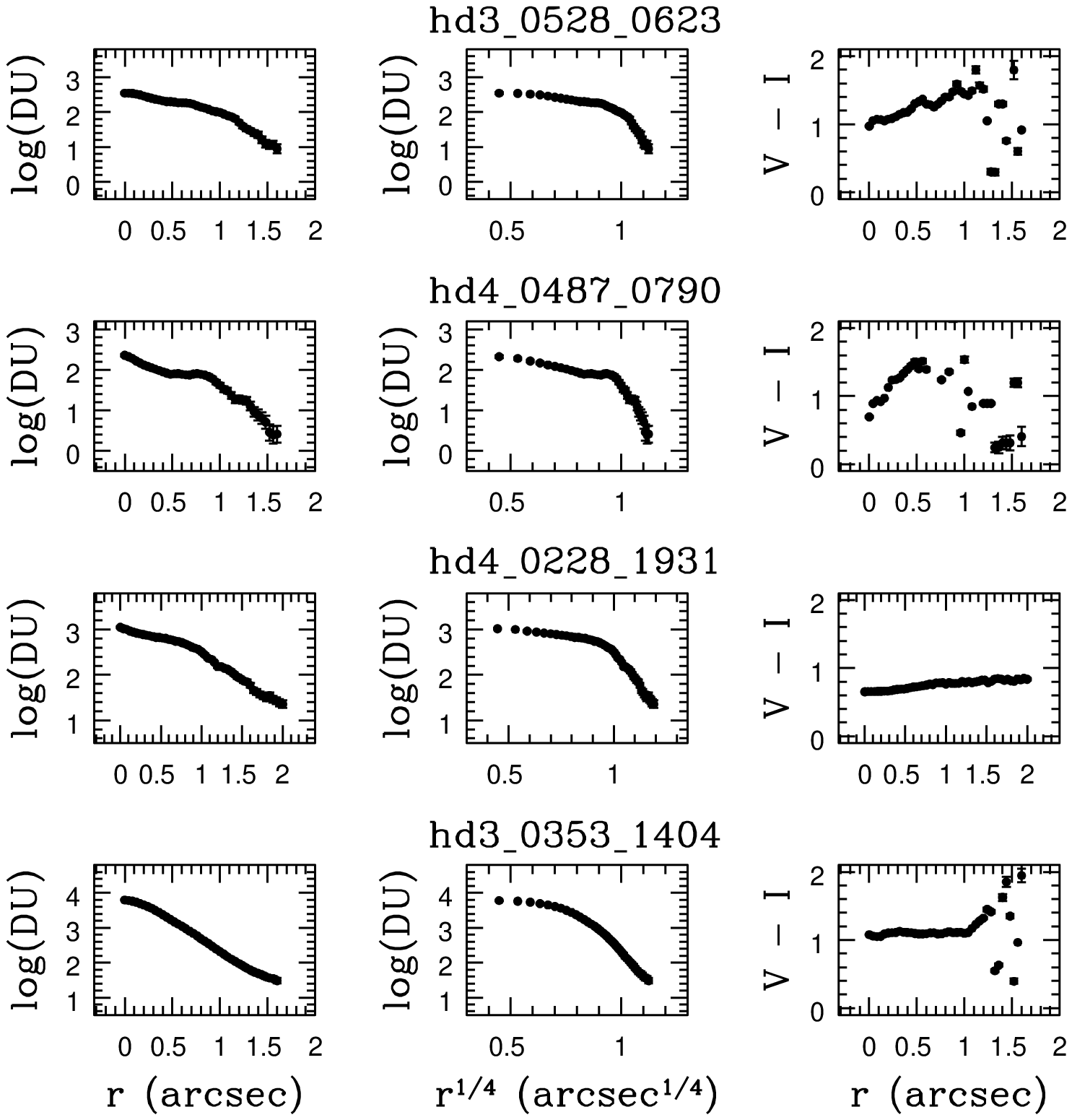]{From top to bottom, examples of photometric 
profiles flatter than a pure exponential for the galaxies 
hd3$_-$0528$_-$0623, hd4$_-$0487$_-$0790, hd4$_-$0228$_-$1931, and 
hd3$_-$0353$_-$1404.  The points are drawn with their 1$\sigma$ 
Poissonian error bars.  The flatness of the profile in the F814W-band 
is due in two cases to a color gradient (top) but in the other two 
cases, the flattening occurs as a real structural property of the 
galaxy (bottom).  \label{colgradexamples}}

\figcaption[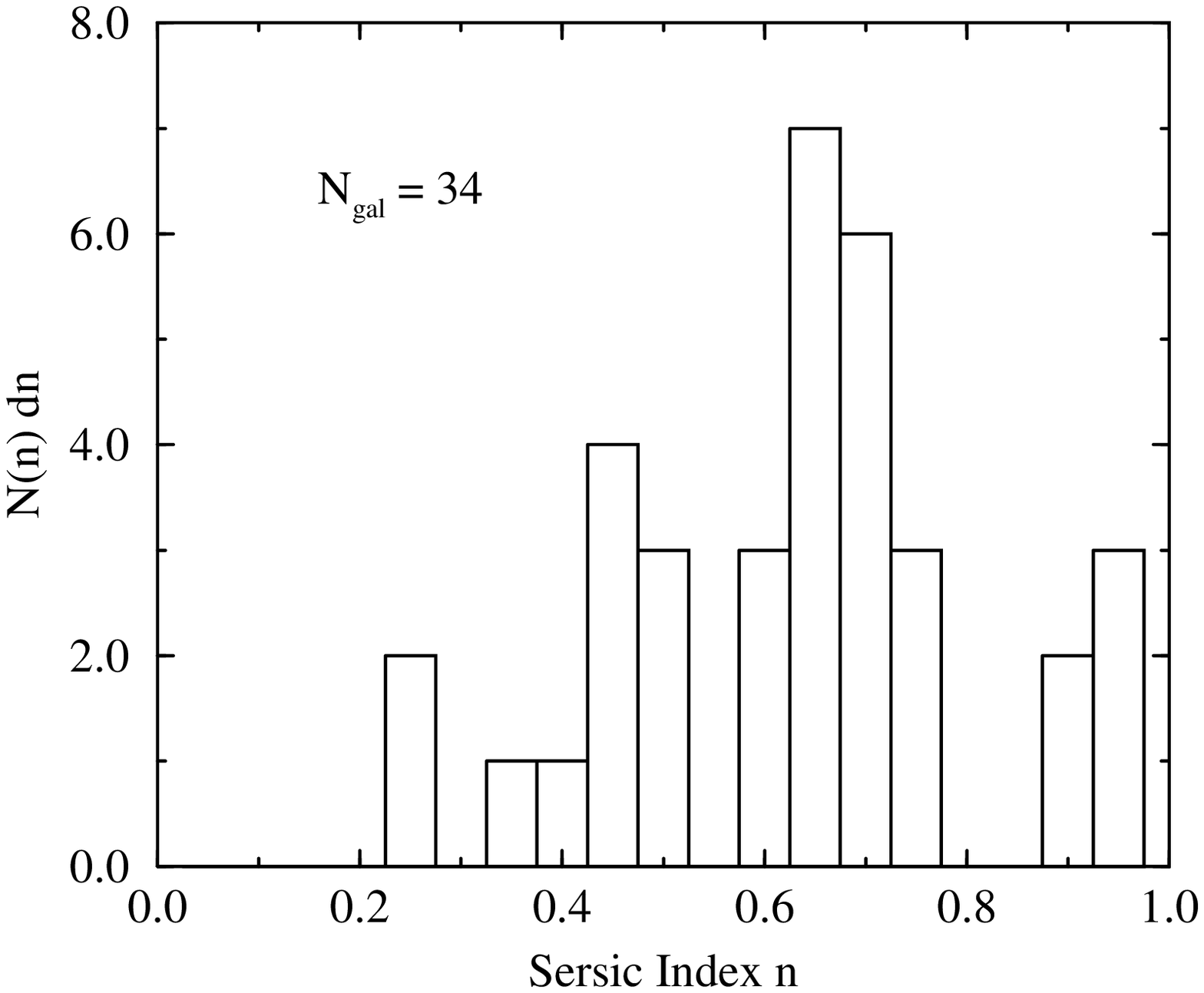]{Distribution of S\'ersic index 
values for S\'ersic galaxies with no color gradient ($\delta 
(V-I)<0.2$).  The median value of $n$ is 0.62, and the dispersion 
$\sigma_{n} = 0.18$.
\label{sersic-n-nocolgrad}}

\figcaption[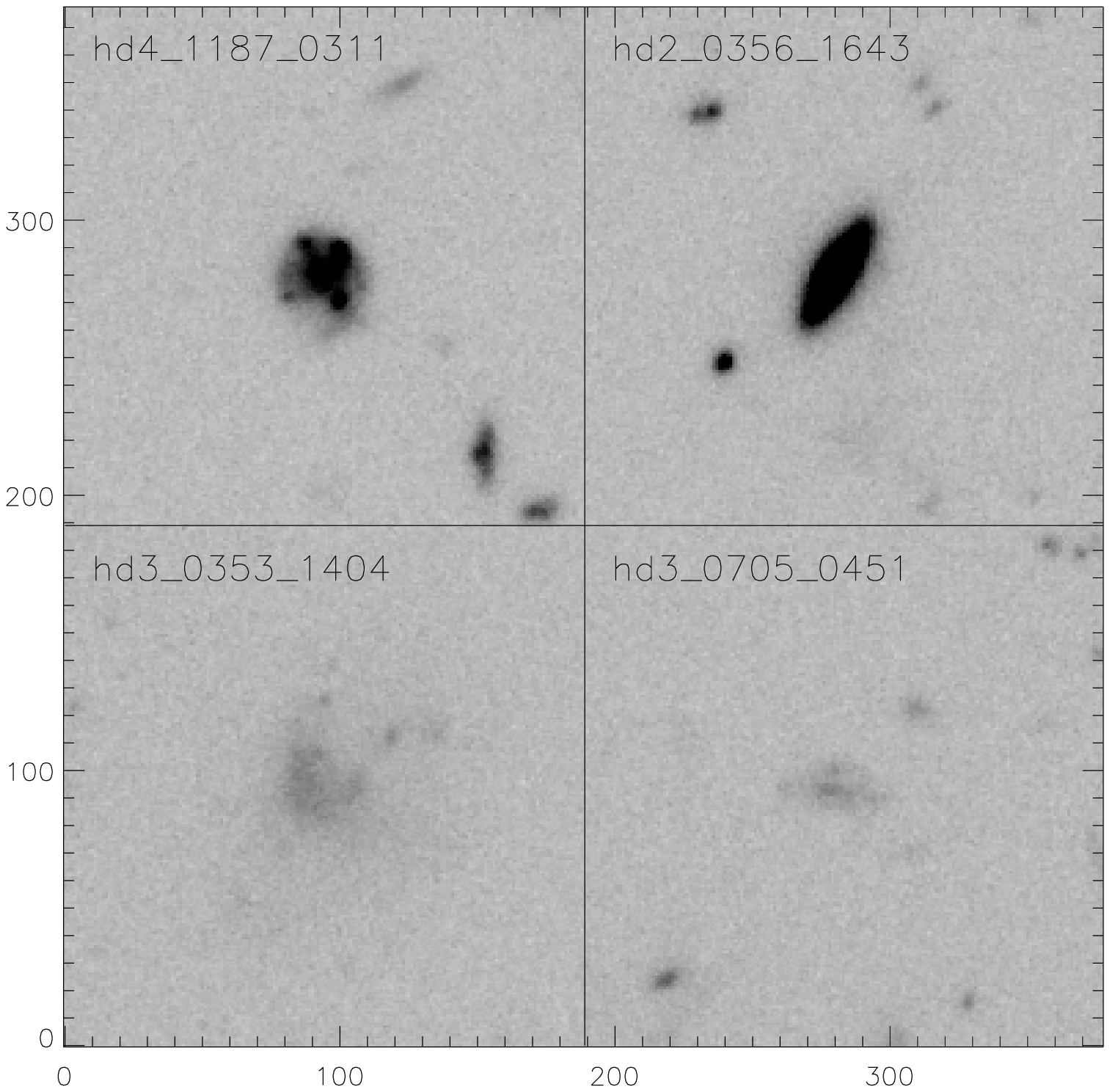]{Mosaic image of four structurally 
anomalous galaxies in the F814W filter.  As these images show, the 
anomalous galaxies appear either as compact looking mergers or more 
diffuse galaxies with surface brightness nebulosities.  The galaxies 
hd4$_-$1187$_-$0311 and hd3$_-$0353$_-$1404, the latter appearing in 
Figure~\ref{colgradexamples}, have no color gradient.  The scale of these images 
is of 7\arcsecpoint 56 on a side.  The objects themselves have 
half-light radii $r_{hl}<$ 0\arcsecpoint 5.
\label{mosaic_sersic}}

\figcaption[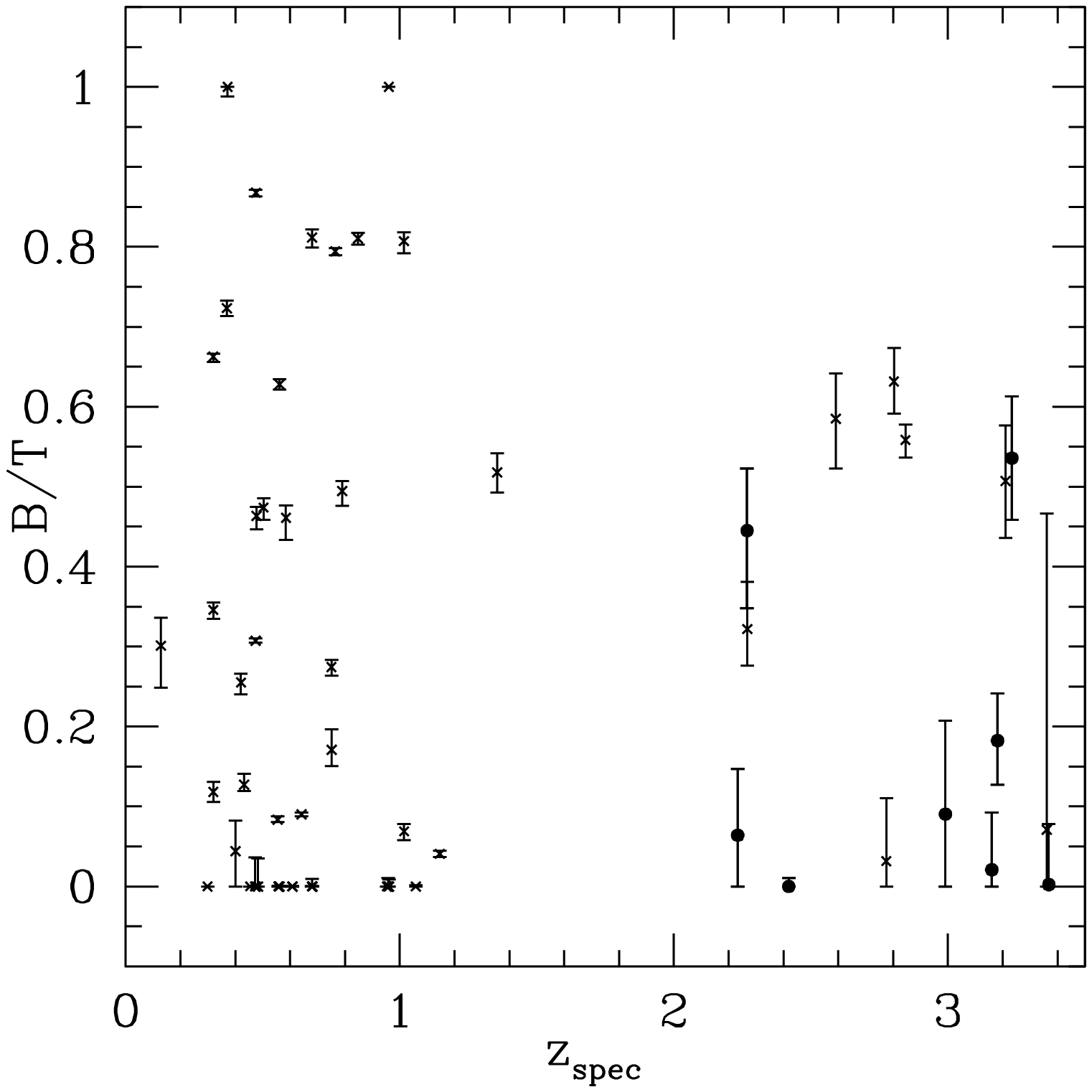]{Bulge fraction as a function of redshift 
for 61 galaxies with published spectroscopic redshifts.  The 99\% 
confidence limits of the bulge fraction parameter are shown with error 
bars.  The points denoted by crosses are the spectroscopic redshifts 
from Cohen \etal (1996), Steidel \etal (1996a), and Zepf \etal (1997) 
and the filled circles are from Lowenthal \etal (1997).  The objects 
selected by Lowenthal \etal (1997) are at $z>2.0$ and have $B/T<0.45$ 
except for one galaxy with $B/T=0.54$.  All the galaxies range in size 
from $0\arcsecpoint 10<r_{hl}<1\arcsecpoint 12$.  The objects with 
$z>2.0$ have half-light radii in the range 2.06-9.28 $h_{50}^{-1}$ 
kpc.
\label{redshift_spec}}

\clearpage
\begin{figure}[h]
\plotfiddle{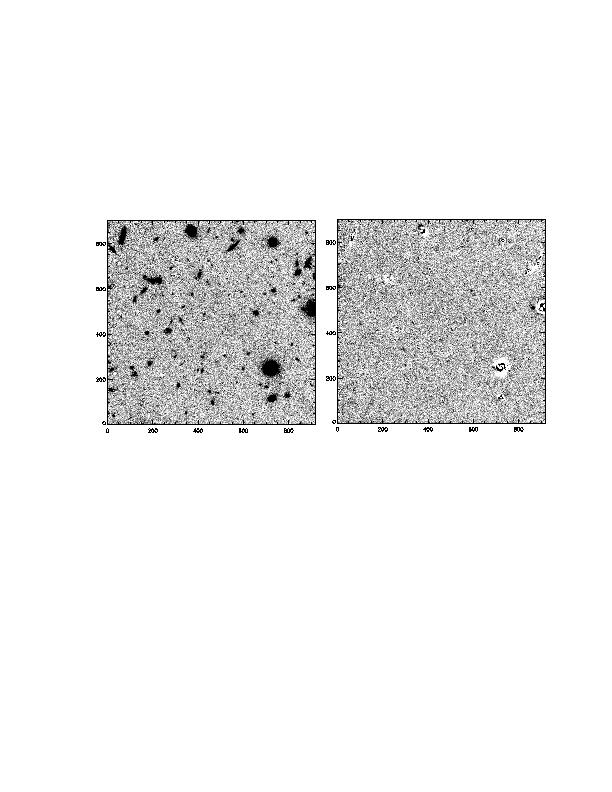}{100pt}{0}{100}{100}{-350}{-570}
\end{figure}

\clearpage
\begin{figure}[h]
\plotfiddle{marleau98_fig2.ps}{200pt}{0}{100}{100}{-290}{-450}
\end{figure}

\clearpage
\begin{figure}[h]
\plotfiddle{marleau98_fig3.ps}{200pt}{0}{100}{100}{-290}{-450}
\end{figure}

\clearpage
\begin{figure}[h]
\plotfiddle{marleau98_fig4.ps}{200pt}{0}{100}{100}{-290}{-450}
\end{figure}

\clearpage
\begin{figure}[h]
\plotfiddle{marleau98_fig5.ps}{200pt}{0}{100}{100}{-300}{-570}
\end{figure}

\clearpage
\begin{figure}[h]
\plotfiddle{marleau98_fig6.ps}{200pt}{0}{100}{100}{-290}{-450}
\end{figure}

\clearpage
\begin{figure}[h]
\plotfiddle{marleau98_fig7.ps}{200pt}{0}{100}{100}{-290}{-450}
\end{figure}

\clearpage
\begin{figure}[h]
\plotfiddle{marleau98_fig8.ps}{200pt}{0}{100}{100}{-290}{-450}
\end{figure}

\clearpage
\begin{figure}[h]
\plotfiddle{marleau98_fig9.ps}{200pt}{0}{100}{100}{-310}{-450}
\end{figure}

\clearpage
\begin{figure}[h]
\plotfiddle{marleau98_fig10.ps}{200pt}{0}{100}{100}{-300}{-450}
\end{figure}

\clearpage
\begin{figure}[h]
\plotfiddle{marleau98_fig11.ps}{200pt}{0}{100}{100}{-320}{-500}
\end{figure}

\clearpage
\begin{figure}[h]
\plotfiddle{marleau98_fig12.ps}{250pt}{0}{100}{100}{-340}{-450}
\end{figure}

\clearpage
\begin{figure}[h]
\plotfiddle{marleau98_fig13.ps}{250pt}{0}{100}{100}{-340}{-450}
\end{figure}

\clearpage
\begin{figure}[h]
\plotfiddle{marleau98_fig14.ps}{250pt}{0}{100}{100}{-340}{-450}
\end{figure}

\clearpage
\begin{figure}[h]
\plotfiddle{marleau98_fig15.ps}{250pt}{0}{100}{100}{-340}{-450}
\end{figure}

\clearpage
\begin{figure}[h]
\plotfiddle{marleau98_fig16.ps}{200pt}{0}{100}{100}{-340}{-500}
\end{figure}

\clearpage
\begin{figure}[h]
\plotfiddle{marleau98_fig17.ps}{200pt}{0}{100}{100}{-340}{-550}
\end{figure}

\clearpage
\begin{figure}[h]
\plotfiddle{marleau98_fig18.ps}{250pt}{0}{100}{100}{-340}{-500}
\end{figure}


\begin{references}
\reference {} Abraham, R.G., Tanvir, N.R., Santiago, B.X., Ellis, 
R.S., Galzebrook, K., \& van den Bergh, S. 1996a, \mnras, 279, L47 (ABR96)
\reference {} Abraham, R.G., van den Bergh, S., Glazebrook, K., 
Ellis, R., Santiago, B.X., Surma, P., \& Griffiths, R.E. 1996b, 
\apjs, 107, 1
\reference {} Abraham, R.G., Valdes, F., Yee, H.K.C., \& van den Bergh, 
S. 1994, \apj, 432, 75 
\reference {} Bertin, E., \& Arnouts, S. 1996, \aaps, 117, 393
\reference {} Bouwens, R.J., Broadhurst, T.J.B., \& Silk, J. 1997, 
astro-ph/9710291
\reference {} Bruzual, A., \& Charlot, S. 1993, \apj, 405, 538
\reference {} Capaccioli, M. 1989, in Corwin H.G., Bottinelli L., 
eds, The World of Galaxies, Springer-Verlag, Berlin, p.208
\reference {} Cohen, J.G., Cowie, L.L., Hogg, D.W., Songaila, A., 
Blandford, R., Hu, E.M. \& Shopbell, P. 1996, \apj, 471, L5
\reference {} Cowie, L.L., Songaila, A., Hu, E.M., \& Cohen, J.G. 
1996, \aj, 112, 839
\reference {} de Vaucouleurs, G. 1948, Ann. d'Ap., 11, 247
\reference {} de Vaucouleurs, G. 1953, \mnras, 113, 134
\reference {} de Vaucouleurs, G. 1956, Occasional Notes RAS, 3, 129
\reference {} de Vaucouleurs, G. 1959, in Handbuch der Physik, 
53, ed. S. Flugge (Berlin:  Springer), 275
\reference {} Ellis, R.S., Colless, M., Broadhurst, T., Heyl, J., \& 
Glazebrook, K.G. 1996, \mnras, 280, 235
\reference {} Giavalisco, M., Livio, M., Bohlin, R.C., Macchetto, 
F.D., \& Stecher, T.P. 1996a, \aj, 112, 369
\reference {} Giavalisco, M., Steidel, C.C., \& Macchetto, F.D. 
1996b, \apj, 470, 189
\reference {} Hubble, E. 1926, \apj, 64, 321
\reference {} Hubble, E. 1930, \apj, 71, 231
\reference {} Kauffmann, G., Charlot, S., \& White, S.D.M. 1996, 
\mnras, 283, L117
\reference {} Krist, J. 1993, in Astronomical Data Analysis 
Software and Systems II, 52, eds. R. J. Hanisch, R. J. V. 
Brissenden, \& Jeannette Barnes (A.S.P. Conference Series), 536
\reference {} Lilly, S.J., Tresse, L., Hammer, F., Crampton, D., \& 
Le F\`evre, O. 1995, \apj, 455, 108
\reference {} Lowenthal, J.D., Koo, D.C., Guzm\'an, R., Gallego, J., 
Philips, A.C., Faber, S.M., Vogt, N.P., Illingworth, G.D., \& 
Gronwall, C. 1997, \apj, 481, 673L
\reference {} Metropolis, N., Rosenbluth, 
A., Rosenbluth, M., Teller, A., and Teller, E. 1953, Journal of 
Chemical Physics, 21, 1087
\reference {} Okamura, S., Watanabe, M., \& Kodaira, K. 1988, in  
The World of Galaxies, ed. H. Corwin \& L. Bottinelli (Berlin:  
Springer), 75
\reference {} Patterson, F.S. 1940, Harvard Bull., 914, 9
\reference {} Sandage, A. 1961, Hubble Atlas of Galaxies 
(Washington:  Carnegie Inst. Washington)
\reference {} Saha, P., \& Williams, T.B. 1994, \aj, 107, 1295
\reference {} Schade, D., Lilly, S.J., Crampton, D., Hammer F., Le 
F\`evre, O., \& Tresse L. 1995, \apj, 451, 1
\reference {} Schade, D., Lilly, S.J., Le F\`evre, O., Hammer, F., \& 
Crampton, D. 1996, \apj, 464, 79
\reference {} S\'ersic, J.L. 1968, Atlas de Galaxias Australes 
(Cordoba:  Observatorio Astronomica)
\reference {} Simard, L. 1998, in preparation
\reference {} Steidel, C.C., Giavalisco, M., Dickinson, M.E., \& Adelberger, K.L. 1996a, \aj, 112, 352
\reference {} Steidel, C.C., Giavalisco, M., Pettini, M., Dickinson, M.E., \& Adelberger, K.L. 1996b, \apj, 462, L17
\reference {} Steinmetz, M., \& M\"uller, E. 1995, \mnras, 276, 549
\reference {} van den Bergh, S., Abraham, R.G., Ellis, R.S., Tanvir, 
N.R., \& Glazebrook, K.G. 1996, \aj, 112, 359 (VDB96)
\reference {} Vanderbilt, D., \& Louie, S.G. 1984, Journal of 
Computational Physics, 56, 259
\reference {} Williams, R.E. \etal 1996, \aj, 112, 1335
\reference {} Zepf, S.E., Moustakas, L.A., \& Davis, M. 1997, \apj, 
474, L1 
\reference {} Zepf, S.E. 1997, astro-ph/9711355 
\end{references}
\end{document}